\documentclass[acmsmall,nonacm]{acmart}
\usepackage{epstopdf}
\usepackage[caption=false]{subfig}

\usepackage{amsmath}
\usepackage{bm}
\usepackage{multirow}
\usepackage{array}
\usepackage{longtable}
\usepackage{tabularx}
\usepackage{booktabs}
\usepackage{adjustbox}
\usepackage{pdflscape}
\usepackage{xtab, afterpage}
\usepackage{threeparttable} 
\usepackage{siunitx}
\usepackage{numprint}

\usepackage{csquotes}
\usepackage{textcomp}
\usepackage{enumitem}
\setlist[itemize]{noitemsep, topsep=0pt, leftmargin=2em}

\usepackage{graphicx} \graphicspath{{./figures/}}
\usepackage{float} 
\usepackage[most]{tcolorbox}
\usepackage[ruled,vlined]{algorithm2e}

\usepackage{cleveref}
\usepackage{todonotes}
\usepackage{soul}  
\sethlcolor{yellow}

\title{Leveraging Mental Models for Human-Centered Explanations: A Systematic Design Approach Illustrated by Explainable Fake Review Identification}

\author{Sidra Naveed}
\authornote{CONTACT Dr. Sidra Naveed. Email: sidra.naveed@uni-siegen.de}
\email{sidra.naveed@uni-siegen.de}
\affiliation{\institution{University of Siegen}\city{Siegen}\country{Germany}}

\author{Md Shajalal}
\affiliation{\institution{University of Siegen}\city{Siegen}\country{Germany}}

\author{Gunnar Stevens}
\affiliation{\institution{University of Siegen}\city{Siegen}\country{Germany}}
\affiliation{\institution{Hochschule Bonn-Rhein-Sieg (H-BRS)}\city{Sankt Augustin}\country{Germany}}

\author{Alexander Boden}
\affiliation{\institution{Hochschule Bonn-Rhein-Sieg (H-BRS)}\city{Sankt Augustin}\country{Germany}}

\begin{document}

\begin{abstract}
Explainable AI (XAI) often prioritizes model fidelity over user intelligibility, yielding explanations that satisfy technical transparency yet fail to support human sense-making. We propose a human-centered approach, \emph{MIND-X} (Mental-model INtegrated Design for eXplainability), which aligns explanations with users’ mental models, commonsense concepts, and judgment heuristics. Using fake review detection as a case study, we contribute: (1) a qualitative study (n=10) that elicits how people assess review credibility and produces an actionable, design-guiding taxonomy of perceptual cues; (2) a set of design principles that link those cues to explanation strategies and prompt LLMs to generate adaptive, language-based explanations; and (3) a quantitative evaluation (n=166) showing that these explanations offer distinct advantages over token-level transparency methods widely used in explainable text classification in terms of understandability, satisfaction, perceived trust, persuasiveness, and decision-making support. Rather than claiming general superiority, our evaluation illustrates how grounding explanations in users’ everyday reasoning practices can produce more intelligible and context-sensitive accounts. Our findings advance human-centered explanation design beyond a mere \textit{``gloss of transparency,''} demonstrating that a mental-model–aligned design process can improve intelligibility and support accountability across user-centered explanation tasks.
\end{abstract}


\maketitle

\section{Introduction}
\label{sec:Introduction}

Artificial Intelligence (AI) systems are becoming increasingly embedded in consequential decision-making contexts, spanning domains such as healthcare, content moderation, education, human resources, and e-commerce.
Within these high-stakes environments, the challenge of \emph{explainability for non-professional users} has emerged as one of the central issues in contemporary human{--}machine interaction~\citep{liao2020questioning,ehsan_accountable_2021}. Explanations for the general public are not merely a matter of technical transparency but must also embody \emph{accountability} and \emph{intelligibility} in ways that are socially recognizable and usable within the everyday reasoning practices of end-users~\citep{mittelstadt_explaining_2019}.

Prior work on explanation design has largely focused on making visible the internal operations of AI systems by exposing input features that are most influential in generating outputs~\citep{bach2024systematic}. In text classification tasks, for instance, methods such as LIME~\citep{ribeiro_lime_2016}, SHAP~\citep{lundberg_shap_2017}, or Layer-wise Relevance Propagation (LRP)~\citep{bach_pixel_2015} highlight specific words or tokens deemed pivotal to a model’s decision. While these approaches are often praised for their \emph{technical faithfulness} to the underlying computation~\citep{doshi2017towards, lipton2018mythos}, they frequently fail to achieve \emph{commonsense meaningfulness} because they do not align with the reasoning practices, interpretive resources, and domain-specific expectations through which people make sense of algorithmic outcomes~\citep{miller2019explanation,ehsan2019automated}. As a result, such explanations risk producing what HCI scholars have criticized as a gloss of transparency—explanations that appear informative but may not reveal the true basis of the system’s decision~\citep{ehsan2019automated},
satisfying methodological rigor while providing limited support for user comprehension, sense-making, or calibrated trust~\citep{abdul2018trends,liao_human-centered_2020}. 

These challenges are particularly evident in the domain of online marketplaces, where customer reviews strongly influence consumer decisions and trust in digital platforms. The rise of generative AI models has made it increasingly easy for malicious actors to produce convincing fake reviews~\citep{cardoso2018towards}. Although state-of-the-art machine learning models can achieve high predictive accuracy in detecting deceptive reviews~\citep{shajalal2025dothe}, they largely function as \textit{``black boxes,''} offering little actionable or user-facing explanations.

In existing explainable text classification systems, explanations are commonly instantiated as word- or token-level importance visualizations (e.g., via LRP). Such approaches have proven useful in domains where class membership is strongly associated with distinctive lexical features or class-specific word sets. However, in the fake review context, both human-written and AI-generated reviews often employ similar vocabularies and surface-level expressions. As a result, highlighted tokens tend to reflect statistical correlations learned by the model rather than cues that users themselves consider meaningful when judging review credibility, limiting their value as sense-making resources for end users.


Platforms such as Amazon Germany provide mechanisms for assessing review authenticity, yet these tools typically do not explain \emph{why} a review is classified as genuine or fake. Without intelligible, context-sensitive explanations, users face uncertainty, which undermines trust and limits accountability~\citep{jacobs2019meaningful,shulner2025made}. 

To address this gap, we advocate for a shift from explanation as the \emph{exposure of internals} to explanation as a \emph{sense-making resource}. We specifically draw on the potential of large language models (LLMs), such as ChatGPT, to serve as \emph{explanation agents} \citep{slack2023explaining}.

Crucially, we do not treat LLMs as generic generators of fluent explanations. Instead, we ground explanation generation in users’ \emph{mental models}—their internal representations of how fake reviews are produced, detected, and interpreted—so that explanations align with how users actually reason about review credibility.

While it may appear straightforward that natural-language explanations generated by large language models (LLMs) would better support user sense-making than word highlighting, language alone does not guarantee intelligibility or relevance. Free-form textual explanations can still emphasize factors that users find unhelpful, overlook critical cues, or over-persuade users without fostering meaningful understanding.

Rather than treating explanation as a one-way projection of system logic, we conceptualize explanations as a process of \emph{account-giving} in the ethnomethodological sense~\citep{garfinkel1967studies}. In this view, explanations function as narratives that are recognizable, adequate, and \textit{``good enough''} within the interpretive horizon of the user. 

This perspective highlights that explanations must be both \emph{domain-oriented} (anchored in the user’s conceptual and practical understanding of the task~\citep{bobek2025user}) and \emph{goal-oriented} (tuned to explanatory objectives such as trustworthiness, simplicity, or persuasiveness~\citep{ehsan_accountable_2021,liao_human-centered_2020}). The task of XAI, then, is not simply to ``open the black box,'' but to provide \emph{accounts} through which users can construct \emph{workable mental models} and engage in practical reasoning~\citep{ehsan_accountable_2021}.

Our human-centered explanation design approach \emph{MIND-X} (Mental-model INtegrated Design for eXplainability) proceeds in six steps: starting from the \textit{AI domain model} (the trained classifier) and the \textit{user domain model} (users’ mental models of the task), a \textit{domain explanation agent} generates user-comprehensible accounts that are then adapted to specific \textit{explanation objectives}. A \textit{user explanation agent} subsequently tailors these explanations to situational needs, producing goal-oriented and cognitively aligned outputs.

To operationalize this perspective, we conducted a structured interview study with ten participants to elicit how people assess review credibility, which cues they attend to, and what they expect explanations to clarify in the fake review context. Analysis of these interviews revealed a set of recurring heuristics and assumptions—such as expectations about experiential detail, narrative coherence, emotional plausibility, and reviewer intent—that together form a user mental model of fake review detection.

The mental models elicited in the interview study are translated into structured prompt templates that guide LLM-generated explanations, ensuring that generated explanations explicitly reference the cues and reasoning strategies users themselves consider meaningful.

To assess whether explanations derived from these user-centered principles are effective in practice, we further conducted a large-scale human-grounded evaluation. In a controlled comparative study with $N=166$ participants recruited via Prolific, we benchmarked language-based explanations generated by large language models (LLMs) against a widely used word-highlighting baseline based on Layer-wise Relevance Propagation (LRP). Participants evaluated both explanation types across six established explanation-quality dimensions, including understandability, satisfaction, trust, decision support, persuasiveness, and perceived cognitive load. This mixed-methods evaluation allows us to systematically examine not only how explanations are designed, but also how they are perceived, interpreted, and used by everyday users in the context of fake review detection.

Finally, the design cycle concludes by grounding the resulting explanations in human-centered evaluation, ensuring that they demonstrably serve explanatory goals through established empirical procedures in XAI research. We demonstrate the implementation of this design approach in the context of fake review detection, which we treat as an illustrative case study rather than a claim of domain-specific superiority. 

\begin{itemize}
 \item We demonstrate how to leverage human-centered design principles to systematically align explanation design with users’ reasoning processes~\citep{norman2013design,muller1993participatory,eiband2018transparency}; 
 \item We demonstrate how empirically elicited user mental models can be operationalized into explanation strategies using LLMs as explanation agents.
\end{itemize}

Specifically, we show how to construct the \textit{user domain model} through a qualitative study, which captures recurring reasoning patterns rather than exhaustively modeling all possible user beliefs, and how to create a structured prompt template~\citep{paraschou2025mind,swamy2025illuminate} that enables LLMs to function as a domain explanation agent.

We further illustrate how explanations can be adapted to predefined \textit{explanation objectives}~\citep{naveed2024overview} through prompt engineering techniques, ensuring that outputs remain responsive to targeted explanation goals and situational demands. Finally, we showcase how to evaluate our approach using a mixed-methods design that combines qualitative insights and quantitative measures~\citep{amershi2019guidelines,liao2020questioning}, benchmarking it against a keyword-highlighting baseline as a representative transparency-oriented technique, rather than as an apples-to-apples competitor to mental-model–aligned explanations~\citep{shajalal2025dothe}.

\section{Related Work}

\subsection{Human-Centered Explanation Design}
\label{sec:explanation-research}

Traditionally, AI systems learn from training data to build models that make predictions or
classifications within a domain. While effective, these models'
outputs often remain opaque, leaving users with the unresolved question of
\textit{``Why?''}. The DARPA XAI program~\citep{gunning2019darpa} responded by
extending the traditional data-to-decision pipeline with an additional explainable
component designed to render system reasoning more transparent.

A broad range of explanation methods have since been developed~\citep{guidotti2018survey,
doshi2017towards}: \textit{feature-attribution techniques} (e.g., LIME, SHAP, LRP~
\citep{kamolov2025feature}), \textit{example-based approaches} (e.g., counterfactuals~
\citep{dwivedi2023explainable}), \textit{concept-based methods} such as TCAV~
\citep{kim2018interpretability}, and \textit{causal representation
learning}~\citep{scholkopf2021toward}. While valuable for experts to debug and
improve model performance, such methods often fail to meet the needs of everyday users, who
require explanations that are both comprehensible and actionable~\citep{schneider2024explainable}.

This limitation stems in part from the fact that most XAI approaches are rooted in a
positivist, causal realist orientation~\citep{woodward2005making}, which conceptualizes
explanations primarily as causal reflections of computational processes rather than as
resources tailored to human reasoning. Their legitimacy is framed in
correspondence-theoretical terms~\citep{tarski1944semantic}, i.e., by the degree to which
they faithfully reflect the AI model, rather than by how they support situated
sense-making~\citep{weick1995sensemaking}.

Beyond technical correctness, prior work demonstrates that explainability fundamentally
shapes users’ perceptions, trust, and acceptance of AI systems. Shin~
\citep{shin2021effects_explainability} shows that explanations affect users not only through
perceived transparency, but also through perceived \emph{causability}—that is, the extent
to which an explanation provides a meaningful causal account of an outcome from the user’s
perspective. These findings suggest that explanations must resonate with users’ causal
reasoning and sense-making practices, rather than merely exposing internal model mechanics.

Lipton~\citep{lipton2018mythos} and Miller~\citep{miller2019explanation} further highlight the
shortcomings of purely model-centric views, arguing instead for explanations grounded in
the social sciences’ understanding of human explanatory practices. This demand echoes a
social-constructivist understanding~\citep{garfinkel1967studies}, in which explanations are conceived as situated, context-sensitive
accomplishments. From this perspective, explanations are understood less as mirrors of a
\textit{ground truth} than as interactional resources accountable to the explainee’s
reasoning, mental models, and objectives.

In response to the shortcomings of existing XAI approaches for everyday users, Abdul et
al.~\citep{abdul2018trends} called for a shift from model-centric fidelity measures to
socio-technical perspectives. In HCI, this epistemic and pragmatic lens emphasizes the
everyday reasoning of stakeholders and the multiple objectives explanations must serve.
Beyond understandability and transparency~\citep{hoffman2018metrics}, the literature
highlights user satisfaction and perceived usefulness~\citep{naveed2024overview},
alignment with mental models~\citep{bansal2019beyond}, persuasive power and decision support~
\citep{vanderwaa2021evaluating}, reduction of cognitive load~
\citep{liao2020questioning}, and effects on trust and confidence~
\citep{kulesza2015principles,amershi2019guidelines}.

Empirical work further demonstrates that trust and reliance emerge from a complex
interaction between model accuracy and explanation form. Papenmeier et al.~
\citep{papenmeier2022s} show that even highly accurate systems can induce over- or
under-reliance when explanations are poorly aligned with users’ expectations, underscoring
the importance of explanation design for calibrated trust rather than blind reliance.

Recent user-centered evaluation research emphasizes that explanation quality must be
assessed \emph{with and for humans}. User-centric evaluation frameworks argue that
explanations should be evaluated in relation to users’ goals, expectations, and
interpretive strategies, rather than relying solely on proxy metrics or expert judgment~
\citep{user_centric_xai_2025}. This perspective directly motivates human-centered,
construct-based evaluation approaches.

At the same time, explanation design must carefully balance informativeness and cognitive
effort. Liedeker et al.~\citep{liedeker2024empirical} identify a ``sweet spot'' of explanation
complexity, showing that overly detailed explanations can overwhelm users, while overly
simplified explanations risk being perceived as unhelpful or misleading. These findings
highlight that explanations must be tailored not only to users’ goals, but also to their
cognitive capacities and task contexts.

To address these needs, explainable systems increasingly draw on user-centered~
\citep{norman2013design} and participatory design traditions~
\citep{muller1993participatory}. Examples include Eiband et al.’s~
\citep{eiband2018transparency} stage-based participatory design process, Liao et al.’s~
\citep{liao2020questioning} ``XAI Question Bank,'' and Chromik and Butz’s~
\citep{chromik2021human} design principles for explanation user interfaces. Complementing
these, Ehsan et al.~\citep{ehsan2019automated} propose natural language rationales as
human-like accounts of system behavior.

Recent work further reframes explainability as an interactive process rather than a static
artifact. Raees et al.~\citep{xu2024interactive_ai} trace a shift from post-hoc explanation
toward interactive human–AI systems, in which explanations adapt to users’ questions,
goals, and contexts over time. This shift emphasizes explanations as dialogical
sense-making resources aligned with users’ evolving mental models.

A promising new direction leverages Large Language Models (LLMs) to generate and evaluate
explanations~\citep{slack2023explaining}. Empirical studies suggest that
users perceive language-based explanations as more intuitive than technical
visualizations~\citep{mitrovic2023chatgpt,huang2024chatgpt}. Our work builds on this line of
research by grounding LLM-generated explanations explicitly in empirically elicited user
mental models, rather than treating LLMs as generic post-hoc narrators.

\subsection{Research on Fake Reviews}
\label{sec:fake-reviews}

Online reviews, whether genuine or AI-generated, are written in natural language and
exhibit high variability in style, tone, and content. Detecting deception in such
unstructured data is a non-trivial task, and current research is dominated by
transformer-based approaches that leverage contextual embeddings and attention mechanisms~
\citep{alqadi2025explainable,mohawesh2024fake,shajalal2024matters}.

Inspired by broader explainability work in text classification, fake review detection has
also integrated post-hoc explanation methods. Techniques such as SHAP, LIME, and Layer-wise
Relevance Propagation (LRP) have been adapted to make model predictions more interpretable~
\citep{shajalal2024matters,alqadi2025explainable}. Yet, most approaches remain model-centric,
focusing on internal attribution rather than aligning with user reasoning processes or
decision needs.

While word- or token-level explanations are effective in domains with class-specific
lexical markers (e.g., sentiment or hate speech classification), fake review detection poses a
fundamentally different challenge. Both deceptive and genuine reviews—whether written by
humans or generated by AI—often share similar vocabularies and sentiment expressions. As a
result, highlighted tokens frequently reflect statistical correlations learned by the
model rather than cues that users themselves associate with review credibility.

Attention mechanisms, sometimes assumed to be inherently interpretable, have yielded mixed results in this context. Mohawesh et al.~\citep{mohawesh2024fake} demonstrated that attention weights can be useful in ensemble models, but their evaluation lacked end-user feedback. Similarly, Mitrović et al.~\citep{mitrovic2023chatgpt} used SHAP to distinguish ChatGPT-generated reviews from genuine ones with high accuracy, yet did not examine how end-users perceive or act upon such explanations. Overall, most approaches remain model-centric, focusing on internal attribution rather than aligning with user reasoning processes or decision needs. 

Behavioral and cognitive research shows that users rarely evaluate reviews as isolated word
sequences. Instead, they rely on higher-level heuristics such as coherence, level of detail,
emotional tone, perceived intent, and contextual cues~\citep{filieri2016makes,
walther2023systematic}.


This misalignment helps explain why purely attribution-based explanations often fail as
sense-making resources in this domain. When explanations emphasize isolated tokens, they
risk contradicting users’ mental models of deception, increasing cognitive load or
undermining trust.

Taken together, these findings motivate explanation approaches that are grounded in users’
mental models, support calibrated trust, and manage cognitive effort. Our work responds to
this gap by first eliciting how users reason about fake review credibility and then using
these empirically grounded mental models to guide the generation and evaluation of
language-based explanations.

\section{Methodological Approach}

\subsection{General Explanation Design Approach}

\begin{figure}[htb]
 \centering
 \includegraphics[width=0.95\textwidth]{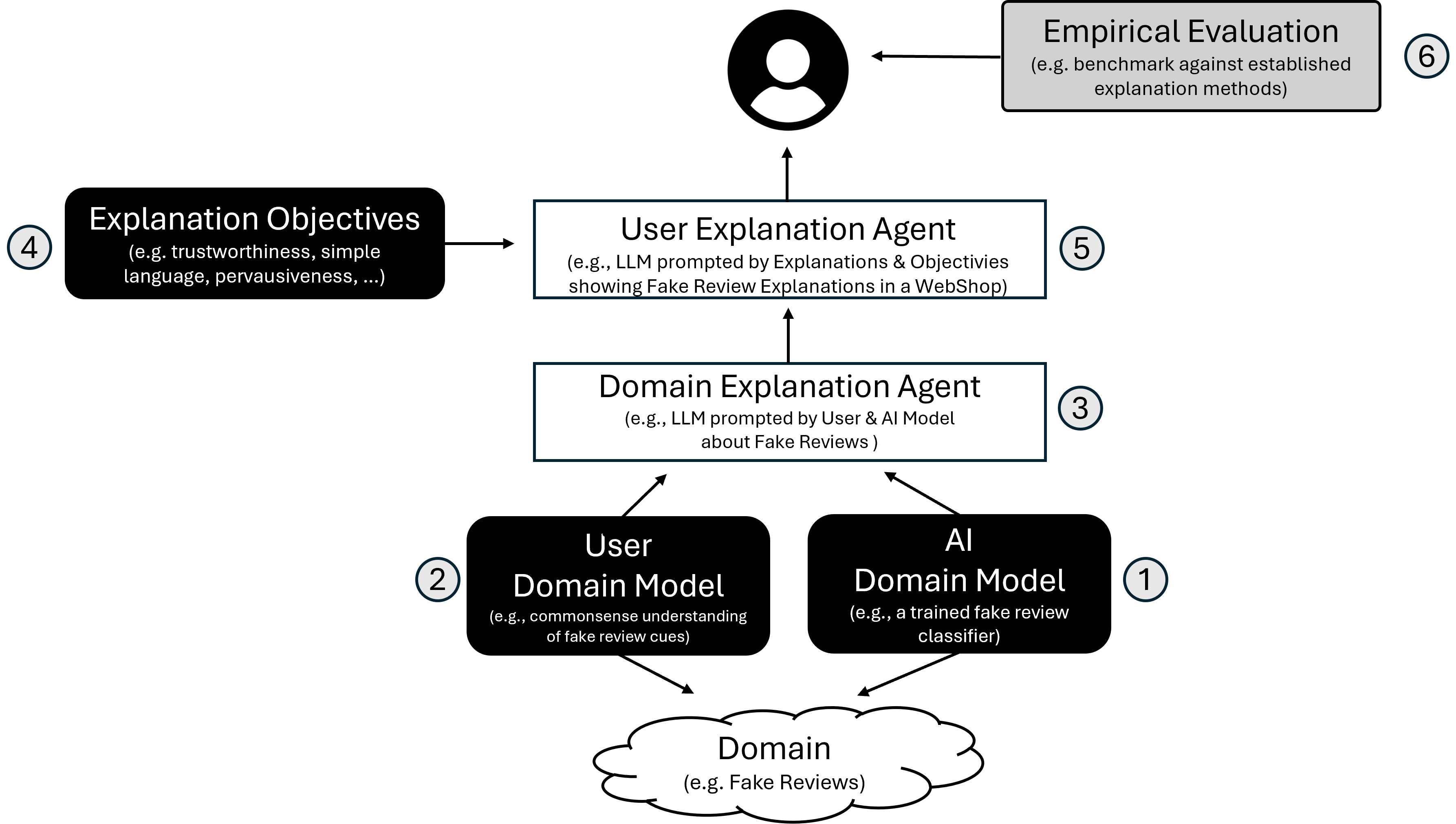}
 \caption{Designing human-centered explanation following MIND-X: From domain models to tailored explanations.
 }
 \label{fig:conceptual-model}
\end{figure}

Addressing the demands outlined in the previous section, our methodological premise is that explanations should be systematically grounded in the mental models of everyday users rather than restricted to the internal mechanics of AI systems. We therefore conceptualize explanation design as a translation process that aligns computational representations with users’ reasoning heuristics and domain-specific understandings. 

Importantly, this premise also responds to recent critiques of explanation paradigms that rely primarily on feature-attribution techniques, which—while effective in certain domains—may fail to support sense-making in tasks where users rely on higher-level semantic and contextual cues rather than isolated features.

Drawing on HCI traditions of human-centered design~\citep{norman2013design}, we propose \emph{MIND-X}, a six-step process consisting of the following elements (see Figure~\ref{fig:conceptual-model}), which we illustrate later in the paper:
\begin{enumerate}
 \item \textbf{AI Domain Model.} Define the trained classifier for the given domain (e.g., a model for detecting fake reviews~\citep{shajalal2025dothe}). This step can be based on traditional AI engineering techniques and common LLM prompting technique.

 \item \textbf{User Domain Model.} Elicit and formalize the corresponding \emph{mental model}~\citep{norman_design_1983} that users hold about the task, derived from prior work or empirical studies.

 \item \textbf{Domain Explanation Agent.} Mediate between AI and user domain models by generating \emph{user-comprehensible accounts}. This step can leverage LLMs through prompt engineering techniques~\citep{white2023promptpatterncatalogenhance}.

 \item \textbf{Define Explanatory Goals.} Make explicit the objectives that explanations should achieve ~\citep{liao2020questioning}. Treat these goals as configurable requirements for tailoring the explanations.

 \item \textbf{User Explanation Agent.} Tailor domain-level explanations to specified goals, situational contexts by adjusting language, framing, and granularity for distinct target groups. This step can also leverage LLMs through prompt engineering techniques~\citep{white2023promptpatterncatalogenhance}.

 \item \textbf{Human-Grounded Evaluation.} Validate the value of the designed explanations in relation to the previously defined explanatory goals. This step should follow the common procedures and best practices in empirical XAI evaluation studies~\citep{naveed2024overview}. 
\end{enumerate}

The main goal of the approach is to extend domain-specific classifiers (\textit{the AI domain model}) with an additional explainable component ~\citep{gunning2019darpa} that aims to explain the particular outcome of the AI model in a manner that users can make sense of. In contrast to approaches that equate explainability with revealing internal model mechanics, MIND-X treats explanations as communicative artifacts whose adequacy must be judged relative to users’ reasoning practices and expectations.
To achieve this, the \textit{user domain model} plays a pivotal role. It captures the \emph{mental models}~\citep{norman_design_1983} that users hold about the task domain, (e.g., commonsense heuristics regarding what makes a review appear fake \citep{walther2023systematic}). 
 
Regarding explanation generation, our method employs a two-step process inspired by the \textit{concept bottleneck principle}~\citep{koh2020concept} as well as the \textit{need-first approach} ~\citep{liao2020questioning}, with exploratory goals addressed as a dedicated step. In the first phase, explanations are aligned with domain-relevant concepts that match user heuristics. In the second phase, these concepts are translated into appropriate linguistic and representational forms, focusing on tone, granularity, and narrative structure. This separation disentangles \textit{what} to explain from \textit{how} to explain, ensuring semantic grounding and communicative adequacy.
Such a design also supports \textit{adaptive and personalized explanations} \citep{Mohammed_2025}, where complexity and detail can be calibrated to user expertise or situational demands. In this way, the \textit{user explanation agent} operationalizes a central HCI principle: aligning transparency with diverse reasoning practices and interpretive resources. The final outcome is explanations that are situated, goal-oriented, and cognitively aligned with the user’s domain model. 

Finally, following recommendations for human-grounded evaluation~\citep{naveed2024overview}, explanations must be validated empirically rather than assessed only in terms of technical fidelity. Beyond model-centric metrics, evaluation should examine whether explanations achieve their specified explanatory goals in real use. This is particularly critical for human-centered approaches grounded in a social-constructivist epistemic stance~\citep{garfinkel1967studies}. From this perspective, formal \textit{model faithfulness} is neither a necessary nor a sufficient condition for effective sense-making~\citep{lipton2018mythos,miller2019explanation}. Instead, the explanatory value must be demonstrated in practice~\citep{naveed2024overview,abdul2018trends}. Consequently, assessing the human-centered qualities of explanations requires established empirical evaluation methods.

\subsection{Instantiation of \emph{MIND-X} for the Design of an Explainable Fake Review Classifier}

The following shows how we apply MIND-X for designing an explainable fake review classifier. This section provides an overview on the key-steps, while details are elaborated in subsequent sections of the paper.

\subsubsection{Step 1: Implementing a Fake Review Classifier}\label{implementing_fake_review}
This step would ordinarily involve the development or fine-tuning of a dedicated fake review classifier, optimized for performance on large-scale benchmark datasets (e.g.,~\citep{shajalal2025dothe,alqadi2025explainable}). Our contribution, however, does not seek to advance classification accuracy but to investigate explanation design. This distinction is critical, as prior work has already established strong baseline performance for fake review detection, the explanatory layer remains underexplored from the perspective of everyday users. For this reason, we deliberately leveraged existing classification capabilities of LLMs (\citep{phoenix2024prompt}), using ChatGPT 5.0 not only to generate human-centered explanation but also to classify the given review simultaneously (as part of Step 3). 

This approach is similar to methods in which models jointly generate a prediction and a natural language explanation ~\citep{narang2020wt5}. As the LLM had correctly classified all reviews used for our human-grounded evaluation (see Step 6), it provided a robust basis for assessing explanation quality without requiring additional model development. Moreover, classification performance in fake review detection is already well-established in prior work~\citep{shajalal2025dothe,alqadi2025explainable}, and re-implementing a state-of-the-art classifier would not have yielded additional insights for our research objectives. 

\subsubsection{Step 2: Eliciting User Domain Models} 
\label{mentalmodel}

The objective of this step was to elicit users' mental models and commonsense concepts \citep{norman_design_1983} related to the heuristics and cues that guide individuals when distinguishing between authentic and deceptive online reviews \citep{walther2023systematic}. Crucially, the prompt does not aim to reproduce the classifier’s internal decision logic, but to reconstruct explanations in a form that mirrors users’ own reasoning strategies identified in Step~2.

To this end, we conducted semi-structured online interviews with 10 participants via Zoom, complemented by a thinking-aloud protocol~\citep{boren2000thinking} in which participants evaluated a set of genuine and fake reviews while verbalizing their reasoning. This exploratory approach allowed us to capture diverse user perspectives and the cues participants rely on when making credibility judgments.
All semi-structured interviews, including the thinking-aloud protocol, were video-recorded, transcribed, and analyzed in MAXQDA\footnote{https://www.maxqda.com/} following the principles of thematic analysis~\citep{braun2012analysis}. Two researchers independently coded each transcript and iteratively refined and consolidated codes into coherent themes through discussion. Coding was primarily inductive, allowing categories and themes to emerge directly from the data. Data collection for the qualitative study was conducted between 13/08/25 and 25/08/25. During this period, semi-structured interviews and thinking-aloud sessions were carried out, transcribed, and subsequently analyzed using thematic analysis. All participants provided written informed consent prior to participation.

Participants were recruited through convenience sampling \citep{etikan2016convenience}, resulting in a group that was diverse in online shopping habits, though uniformly well-educated (all held at least a university degree; see \autoref{fig:demographics}). While this recruitment strategy limits statistical generalizability, it was sufficient for our purposes: the resulting user domain models directly informed the design of our explanations, which later proved to be of high quality in the evaluation study. This outcome suggests that even with a small, non-representative sample, the elicited heuristics captured essential patterns of user reasoning relevant to explanation design.

\begin{table}[!htb!]
 \centering
 \caption{Demographics of participants and their personal experiences with online shopping and reviews}
 \label{fig:demographics}
 \renewcommand{\arraystretch}{1.22}
 \resizebox{\textwidth}{!}{%
 \begin{tabular}{lcccc}
 \toprule
 \textbf{Participant} & \textbf{Occupation} & \textbf{Field of Study} & \textbf{Frequency of Online Purchase} & \textbf{Frequency of Reading Online Reviews} \\
 \midrule
 P1 & Doctoral Student & HCI & Monthly & Often for expensive items; Rarely for low-cost \\
 P2 & Trainee Engineer & Computer Science & Every 1--2 months & Always before purchase \\
 P3 & Master’s Student & HCI & As needed & Always before purchase \\
 P4 & Researcher & Global Politics & 1--2/week & Always before purchase \\
 P5 & Master’s Student & Visual Computing & 2--3/month & Often for all items \\
 P6 & Postdoc Researcher & Political Economy & 1--2/month & Always before purchase \\
 P7 & Master’s Student & AI/ML & Every 2--3 months & Often for expensive items; Rarely for low-cost \\
 P8 & Doctoral Student & Entrepreneurship & Almost daily & Often for technical items; Rarely for clothing \\
 P9 & Master’s Student & Computer Science & 5--6/month & Often for expensive items; Sometimes for small \\
 P10 & Senior Researcher & HCI & Very often & Rarely for all items \\
 \bottomrule
 \end{tabular}
 }
\end{table}

\subsubsection{Step 3: Translating User's Domain Models into Prompt-Templates} 
\label{translation-of-domain-models}
In this step, the objective is to implement a \textit{domain explanation agent}. This can be achieved either by developing a dedicated AI model or by adapting and fine-tuning existing ones. In our work, we chose to realize the agent through prompt engineering~\citep{phoenix2024prompt}, leveraging the capabilities of state-of-the-art LLMs to generate user-centered explanations. 

Building on the thematic analysis, the elicited cues and heuristics were formalized into a structured meta-prompt that guided the LLM’s reasoning. Rather than directly mirroring the internal mechanisms of the model, the prompt was designed to align with users’ cognitive heuristics by embedding cues (e.g., linguistic inconsistency, emotional tone, reviewer credibility) into an explicit reasoning schema. Following prior work on prompt engineering for alignment ~\citep{white2023promptpatterncatalogenhance}, we operationalized a two-stage structure: (1) \textit{Classification} (real vs.\ fake) and (2) \textit{Human-like Reasoning}, expressed through a standardized ``Cue $\rightarrow$ Perception $\rightarrow$ Explanation'' sequence. This design ensured that each identified signal was not only detected but also contextualized in terms of how a human user might perceive and interpret it. To foster naturalistic communication, outputs were constrained to everyday language, with balanced pros and cons, as well as conflict resolution when cues pointed in opposing directions. The resulting prompt-template had modular fields (e.g., \textit{[decision outcome]} + \textit{[user-relevant cue]} + \textit{[justification]}), producing structured yet human-centered narratives such as: "This review is likely fake because [\textit{cue: linguistic inconsistency}] and [\textit{cue: reviewer credibility issue}]."

Templates were iteratively refined through pilot testing, during which two researchers independently reviewed and assessed the quality of generated explanations. The procedure was repeated until both researchers agreed that the explanations were sufficiently clear, consistent, and aligned with user-relevant cues. Explanations were generated using ChatGPT~5.0. For both the pilot phase and the subsequent evaluation study, reviews were drawn using a stratified random, balanced sample of fake and genuine reviews taken from the open dataset provided by \citep{salminen2022creating}. 

\subsubsection{Step 4: Specify Explanation Objectives}\label{specify_explanation}
In this step, the task is to make explicit which objectives the explanations are intended to serve. Prior work has shown that explanation goals often remain implicit, leading to a mismatch between what systems provide and what users actually need. Yet, needs-first approaches, such as the XAI Question Bank~\citep{liao2020questioning}, highlight the value of articulating explanatory purposes explicitly, so that design can be systematically aligned with users’ reasoning practices. 
Because our study was not situated in a single, concrete application context (e.g., being embedded in a deployed review platform) but rather aimed at conducting a generalizable evaluation of explanation design, we sought to incorporate a broad set of objectives identified in the literature. Specifically, we considered six explanation dimensions that are common in XAI research~\citep{naveed2024overview, hoffman2018metrics}: \textit{understandability}, \textit{satisfaction}, \textit{trust}, \textit{cognitive load}, \textit{decision support}, and \textit{persuasiveness}.

Making these objectives explicit avoids post-hoc rationalization and ensures that explanation quality is assessed relative to clearly articulated user-centered criteria rather than abstract notions of interpretability. Together, these dimensions capture both \emph{subject-oriented measures} (e.g., understandability, satisfaction) and \emph{action-oriented measures} (e.g., decision support, persuasiveness), which are particularly salient in light of the threat of consumer manipulation through fake reviews~\citep{walther2023systematic}.

\subsubsection{Step 5: Tailoring Explanation to Explanation Goals and Use Context}
\label{sec:tailoring}
This step involves tailoring domain-based explanations to the defined explanatory objectives, the specific application context, and the intended target audience. This step resembles the use of LLMs as an intermediate layer between the explanation model and the explanation interface~\citep{zytek2024explingo, qiu2024scenarios}, 
where model outputs are adapted into natural language explanations tailored to specific user roles and application contexts. 

In this context, word-highlighting explanations—which are effective in domains with stable class-specific lexical markers—are less suitable because both genuine and deceptive reviews often employ similar vocabulary, requiring higher-level semantic reasoning.

As fake reviews represent one of the most prevalent threats to online marketplaces such as Amazon, our design emphasized two contextual requirements. First, the \textit{target group} consists of everyday web shop users without technical expertise, who require explanations in plain, non-technical language that resonate with familiar heuristics, such as linguistic style, level of detail, or reviewer credibility cues~\citep{filieri2016makes, walther2023systematic}. Second, the \textit{use context} is characterized by consumer decision-making under time pressure, where users must make rapid judgments about credibility. 

Based on this contextual specification, we realized this step by leveraging the capabilities of ChatGPT 5.0 through prompt engineering. To this end, we iteratively developed a prompt template that adapts explanations to contextual demands. By adopting the concept bottleneck principle \citep{koh2020concept}, we passed the output from Step 3 (the \textit{domain explanation agent}) only, rather than the original review text. This design choice ensured greater control over the generation process and helped align the produced explanations with the specified user domain model.
The resulting meta-prompt was explicitly framed to generate explanations that are \emph{easy to understand, trustworthy, and actionable} for everyday web shop users, who must rapidly assess review credibility in order to make informed purchasing decisions. 
 
\subsubsection{Step 6: Human-Grounded Evaluation}\label{human_grounded_evaluation}
The final step was to conduct an empirical evaluation study to assess whether the explanatory goals defined in Step~3 were achieved. Following established practices of human-grounded evaluation in XAI~\citep{naveed2024overview,doshi2017towards}, we specified the target group, operationalized the evaluation objectives, and conducted a benchmarking study using a mixed-methods approach. The quantitative evaluation study was conducted between 30/08/2025 and 15/08/2025 using the Prolific platform. All participants provided informed consent electronically before accessing the survey.

A total of $n = 166$ participants were recruited via Prolific (192 prior to data cleaning), serving as proxy users for the target group of online shoppers. Eligibility criteria required fluency in English and prior experience with online shopping. No further demographic restrictions were applied to capture the diversity of everyday consumers. 
The final sample included 96 participants identifying as female (58\%) and 70 as male (42\%); none identified as non-binary nor preferred not to disclose their gender. The mean age was 39.2 years ($SD = 12.7$), ranging from 20 to 75 years. 

Regarding online shopping practices, nearly all participants reported consulting reviews before purchases, with 154 (93\%) doing so frequently and 12 (7\%) doing so only occasionally. When asked about their level of suspicion regarding fake reviews, 9 participants (5\%) reported never or very rarely experiencing such suspicion, 91 (55\%) reported sometimes suspecting reviews, and 66 (40\%) reported doing so often. 
In terms of familiarity with Artificial Intelligence (AI), 2 participants (1\%) reported low familiarity, 93 (56\%) moderate familiarity, and 71 (43\%) high familiarity. 
Table~\ref{tab:sample} summarizes the demographic and review-related characteristics of the sample.

\begin{table}[!htb]
\centering
\caption{Demographic and review characteristics of the Prolific sample ($N = 166$).}
\label{tab:sample}
\renewcommand{\arraystretch}{1.2}
\begin{tabular}{p{6cm}p{3cm}p{3cm}}
\toprule
\textbf{Characteristic} & \textbf{$n$} & \textbf{\%} \\
\midrule
\textit{Gender} & & \\
Female & 96 & 58 \\
Male & 70 & 42 \\
Non-binary / Prefer not to disclose & 0 & 0 \\
\midrule
\textit{Age} & & \\
$M$ = 39.2, $SD$ = 12.7, Range = 20.0--75.0 & & \\
\midrule
\textit{Review consultation frequency} & & \\
Frequently (before most purchases) & 154 & 93 \\
Occasionally & 12 & 7 \\
\midrule
\textit{Suspicion reviews might be fake} & & \\
Never/Very rarely & 9 & 5 \\
Sometimes & 91 & 55 \\
Often & 66 & 40 \\
\midrule
\textit{Familiarity with Artificial Intelligence (AI)} & & \\
Low familiarity & 2 & 1 \\
Moderate familiarity & 93 & 56 \\
High familiarity & 71 & 43 \\
\bottomrule
\end{tabular}
\end{table}

The study employed a within-subjects A/B design with two conditions: a baseline condition using word highlighting and a treatment condition using LLM-based narrative explanations, which were generated through our human-centered methodology. We selected \textit{word highlighting} as the baseline because it represents a widely used explanation approach in fake review detection research~\citep{kim-etal-2021-distinguish,shajalal2025dothe}. Specifically, we employed the fake review classifier and the LRP-based word-highlighting explanation model described by~\citep{shajalal2025dothe} as our benchmark. 
Each dimension was measured through multiple items adapted to the comparative format (see Table~\ref{tab:constructs}). For every item, participants indicated their judgment using a five-point forced-choice differential scale ranging from \textit{“Much better Language-based”} to \textit{“Much better Word-highlighting”}. This comparative design required participants to directly contrast the two explanatory paradigms rather than providing absolute ratings, thereby yielding a more robust assessment of the relative explanation quality.

The comparison is not intended to evaluate which explanation type is universally superior, but to assess their relative effectiveness within the specific cognitive and contextual demands of fake review detection.

\begin{table}[!htb]
\centering
\caption{Operationalization of explanation quality constructs through comparative questionnaire items.}
\label{tab:constructs}
\renewcommand{\arraystretch}{1.2}
\resizebox{\textwidth}{!}{%
\begin{tabular}{p{3cm} p{10cm} p{3cm}}
\toprule
\textbf{Construct} & \textbf{Items (comparative wording)} & \textbf{Reference} \\
\midrule
Understandability & 
\begin{itemize}[leftmargin=*]
 \item I found the explanation easier to understand. 
 \item The explanation was clearer and less ambiguous. 
 \item The explanation made the AI’s decision-making process more transparent. 
\end{itemize} 
& \citep{hoffman2018metrics,naveed2024overview} \\
\midrule
Satisfaction & 
\begin{itemize}[leftmargin=*]
 \item I am more satisfied with the explanation. 
 \item The explanation met my expectations better. 
\end{itemize} 
& \citep{hoffman2018metrics} \\
\midrule
Trust & 
\begin{itemize}[leftmargin=*]
 \item The explanation increased my trust in the system’s decision. 
 \item The explanation made me believe the AI made the right decision. 
\end{itemize} 
& \citep{amershi2019guidelines} \\
\midrule
Decision Support & 
\begin{itemize}[leftmargin=*]
 \item The explanation helped me make a better decision. 
 \item The explanation made me feel more confident in classifying the review. 
\end{itemize} 
& \citep{naveed2024overview,kulesza2015principles} \\
\midrule
Persuasiveness & 
\begin{itemize}[leftmargin=*]
 \item The explanation convinced me more that the review was (not) fake. 
 \item The explanation influenced my opinion about the review more. 
\end{itemize} 
& \citep{kulesza2015principles,amershi2019guidelines} \\
\midrule
Cognitive Load & 
\begin{itemize}[leftmargin=*]
 \item It was more mentally demanding to understand the explanation. 
 \item I had to put more effort into making sense of the explanation. 
\end{itemize} 
& \citep{naveed2024overview} \\

\bottomrule
\end{tabular}%
}
\end{table}

In addition to the comparative ratings, we included a set of open-ended questions, as these allowed participants to articulate subjective perceptions, unexpected concerns, and contextual nuances~\citep{nielsen1993usability}. Participants were asked to reflect on (a) their overall impression when comparing the two explanation types. (b) in what ways the language-based explanation was better than the word-highlighting explanation and (c) visa versa, (d) potential improvements to the language-based explanation, and (e) word-highlighting explanation.

The survey procedure unfolded in three stages. First, participants completed socio-demographic questions. Second, they proceeded through six evaluation rounds. In each round, the scenario along with the corresponding product review was presented alongside both explanation conditions (see Figure~\ref{fig:interface} and Appendix~\ref{appendix}, which presents the complete explanation examples for all six evaluation scenarios (Figures~\ref{fig:R1Explanations}--\ref{fig:R6Explanations}).), which participants then evaluated across the predefined quality dimensions using the comparative questionnaire. To ensure sufficient engagement, a timer enforced a minimum of two minutes per round. This duration was chosen to balance data quality with ecological validity, as prior research demonstrates that users in everyday online shopping rarely engage in deep analytic reasoning but rely on heuristics to make rapid credibility judgments~\citep{walther2023systematic}. We also included an attention-check item in each round to ensure that participants completed the questionnaire carefully. 

Third, after completing the comparative evaluations, participants responded to a set of open-ended questions to give additional qualitative feedback which cannot captured through closed-ended scales~\citep{nielsen1993usability}. We asked participants to give feedback on (a) their overall impression when comparing the two explanation types, (b) in what ways the language-based explanation was better than the word-highlighting explanation, and (c) vice versa. Further we asked about potential improvements for (d) the language-based explanation, and (e) the word-highlighting explanation. This qualitative feedback complements the insights gained from statistical methods.

All data were screened for completeness and consistency prior to analysis. 
From the initial sample of 192 participants, 26 were excluded because they did not pass all attention checks, resulting in a final sample of $N = 166$. 
The statistical analysis included a psychometric examination of construct validity as well as an adjustment of the significance level to account for multiple $t$-tests (see Section \ref{sec:Validity}). All statistical analyses were conducted in \textit{R} (Version 4.5.1).

The qualitative responses to the open-ended questions were analyzed using thematic analysis, following the same methodological principles applied in Step~2. 
This mixed-method analysis enabled us to combine statistically robust findings with qualitative insights, offering a richer understanding of how users perceived and made sense of the explanations. This triangulation was critical for validating the explanatory goals and assessing their alignment with users’ reasoning practices in the context of fake review detection. \\

\noindent\textbf{\textit{Ethical Statement}}
The study was conducted in accordance with the ethical standards of the University of Siegen and the principles of the 1964 Helsinki Declaration and its subsequent amendments. Based on a preliminary review of the project-related issues, the Ethics Council of the University of Siegen determined that no formal ethical approval was required for this study (Reference No.~LS\_ER\_34\_2026).


For both the qualitative and quantitative studies, all participants provided written
informed consent prior to participation. Participation was voluntary, and participants could withdraw at any time without penalty. All personally identifying information in the transcripts was removed to ensure anonymity. No sensitive personal information was collected or analyzed. Participants received appropriate compensation for their involvement. All procedures for data collection and analysis were carried out in compliance with institutional ethical standards governing research with human participants.


\section{Eliciting and Operationalizing Users’ Domain Model}\label{sec:evaluation}
In this section, we elaborate on the implementation of Steps 2, 3, and 4 of the \emph{MIND-X} framework, as introduced in Section~\ref{mentalmodel}, Section~\ref{translation-of-domain-models}, and Section~\ref{specify_explanation}, respectively. We focus on the elicitation of users’ domain models, their subsequent operationalization into explanation mechanisms, and the adaptation of these mechanisms to defined explanation objectives.

\subsection{Users’ Perceptions of What Makes a Review ``Fake''}

As part of our methodological approach, uncovering shared mental models, salient cues, and heuristics is pivotal for designing explanations that align with human reasoning. Accordingly, we discuss the findings of our qualitative study in depth in the remainder of this section. 

This step directly addresses a key limitation of commonly used explanation techniques in fake review detection. While word- or token-level highlighting can be informative in text classification tasks with class-specific lexical patterns, prior work and our reviewer feedback highlight that, in the fake review domain, both human-written and AI-generated reviews frequently employ similar vocabulary. As a result, isolated word attributions often fail to map onto the cues that users themselves rely on when assessing credibility. Eliciting users’ mental models therefore serves as a necessary methodological step to identify which signals are actually meaningful for human sense-making in this domain.

With regard to mental models of what constitutes a ``fake'' review, participants described fakes as communicative acts with ulterior motives: texts written not to share experiences but to persuade, promote, or deceive. Fake reviews were imagined as strategically constructed performances that exaggerate benefits, suppress flaws, or mimic marketing language. Alongside these broader perceptions, the qualitative analysis reveals how participants oriented themselves toward a set of recurrent cues and cognitive heuristics when evaluating whether a review was authentic or fabricated (see Table~\ref{tab:thematic-summary}). These cues were not experienced as isolated markers, but as interpretive resources that participants mobilized in flexible and sometimes conflicting ways.

\begin{table}[!htb]
 \caption{Summary of cues participants used to evaluate reviews as fake}
 \label{tab:thematic-summary}
 \renewcommand{\arraystretch}{1.2}
 \begin{tabular}{p{4cm} p{4cm} p{6cm}}
 \toprule
 \textbf{Theme} & \textbf{Cue}& \textbf{Description}\\
 \midrule
 Simple Language and Conciseness & Plain, everyday words; short reviews & Signals authenticity; resembles ordinary consumer voice \\
 Excessive Detail & Overly long or elaborate descriptions & Triggers suspicion; feels like product description \\
 Balanced Pros and Cons & Mentions both strengths and weaknesses & Signals honesty and credibility \\
 Emotional Authenticity and Redundancy & Genuine joy; repeated phrases; exaggerated positivity & Natural emotion increases authenticity; over-positivity raises doubt \\
 Persuasive Intent & Repetitive recommendation; promotional tone & Feels promotional; undermines trust \\
 Visual Presentation & Everyday, casual photos vs. polished, staged photos & Casual photos signal authenticity; staged images trigger suspicion \\
 Logical Consistency and Relevance & Contradictions; irrelevant storytelling & Inconsistency or digression undermines credibility \\
 Reflections on Misclassification & False positives, false negatives; heuristic re-evaluation & Awareness of fallibility; adaptation to adversarial cues \\
 \bottomrule
 \end{tabular}
 \end{table}

A first prominent theme concerned the role of \textit{simple language and conciseness}. Participants frequently associated plain, everyday wording and short length with authenticity, describing such reviews as casual and straightforward. As one noted: \emph{“Most real reviews are in very short, simple words like ‘good quality, fine with me’”} (P5). Concise accounts were treated as honest and aligned with expectations of how an ordinary consumer would write.

In contrast, \textit{excessive detail} was treated as a red flag. Long, elaborate reviews were often read as overcompensating or resembling product descriptions rather than personal experience. One participant captured this view directly: \emph{“If the review is too long, that makes me doubt”} (P2). The length and density of description thus functioned as \textit{triggers of suspicion}. 
Another central heuristic was the presence of \textit{balanced pros and cons}. Reviews that acknowledged both strengths and weaknesses were considered particularly trustworthy. For example, P10 observed: \emph{“They mentioned the fabric was pleasant but slippery — that balance convinces me”}. This capacity to highlight both positive and negative features was consistently interpreted as a signal of authenticity. 

Participants also highlighted the importance of \textit{logical consistency and relevance}. Reviews that included contradictions or strayed into irrelevant storytelling were quickly marked as suspicious. For example, P6 emphasized: \emph{“Contradictions in the story make me suspicious”}, while P10 noted: \emph{“If they talk about random things not about the product, I mark it fake”}. Coherence and product focus were thus treated as essential conditions for credibility. 

\textit{Emotional authenticity} further shaped credibility judgments, but in ambivalent ways. Genuine expressions of joy or enthusiasm reinforced trustworthiness, with some participants even interpreting redundancy in wording as a marker of real excitement: \emph{“When someone is too excited, repeating the same word, I think it’s real emotion”} (P6). At the same time, exaggerated positivity or excessive optimism provoked skepticism: \emph{“Optimistic and too much positive… makes me suspicious”} (P6). In some cases, participants admitted that affective expressions could override rational assessment: \emph{“Sometimes the excitement makes me ignore logic and think it’s genuine”} (P2).
Closely linked was a sensitivity to \textit{persuasive intent}. Participants distrusted reviews that seemed to push sales, particularly when recommendation phrases or superlatives were repeated. As P2 put it: \emph{“Every 1 or 2 lines they are recommending to buy this product”}. Such overtly promotional tones were interpreted as contrived and undermined trust. 

Beyond text, \textit{visual presentation} emerged as another salient cue. Everyday, casual photos were described as evidence of genuine product use, whereas polished, studio-like images raised suspicion. As P1 explained: \emph{“High quality, studio-like photos… I trust them less, seems like paid review”}. Thus, photographic style provided an additional dimension for assessing authenticity.

Finally, participants reflected on the limits of their heuristics through experiences of misclassification. \textit{False positives} occurred when polished or detailed reviews were incorrectly assumed fake, while \textit{false negatives} arose when simple wording disguised inauthentic content. As P1 admitted: \emph{“Plain English, repeated words... I thought no one would pay for this. So it felt real”}. Exposure to system classifications sometimes prompted a re-evaluation of assumptions: \emph{“I used to believe fake reviews are long and polished, but as it turns out, not necessarily”} (P1). Others highlighted the adaptive strategies of fake reviewers: \emph{“It has become more difficult to identify fake reviews, they adapt”} (P4). These reflections underscore participants’ awareness that cues can misfire and that authenticity judgments require weighing multiple, sometimes conflicting signals.

Importantly, these findings should not be understood as a descriptive account of user behavior alone, but as a design-relevant abstraction of how people operationalize credibility judgments in practice. The identified cues and heuristics form the conceptual foundation for explanation design, specifying what users expect explanations to reference in order to be perceived as intelligible and trustworthy. This abstraction step is critical for translating qualitative insights into explainable system behavior.

Taken together, the analysis demonstrates that judgments of review authenticity were anchored in a repertoire of recognizable cues: \textit{simple language and conciseness, excessive detail, balanced pros/cons, emotional tone and redundancy, persuasive intent, visual presentation, logical consistency and relevance}, as well as an evolving awareness of \textit{heuristic fallibility}. These themes highlight how authenticity assessments are interactional accomplishments shaped by observable features, affective responses, and adaptive reasoning.

\subsection{Operationalizing Domain-Sensitive Explanation Agents}
\begin{figure}[!htb]
 \centering
 \begin{tcolorbox}[colback=gray!10, colframe=gray!40, width=0.95\textwidth, boxrule=0pt, arc=2pt, left=4pt, right=4pt]
 \small
 \textbf{Meta-Prompt for Review Authenticity Analysis} \\[0.5em]
 
 \textbf{Input:} Product review text $R$ \\[0.3em]
 \textbf{Output:} Classification (\texttt{Real/Fake}) + Human-like reasoning with Cue $\rightarrow$ Perception $\rightarrow$ Explanation chains
 \vspace{0.7em}
 
 \textbf{Stage 1: Classification} \\
 \hspace*{1em} Read the review $R$ carefully. \\
 \hspace*{1em} Decide whether $R$ is more likely \texttt{Real} or \texttt{Fake}. \\
 \hspace*{1em} Base the judgment on cues such as language style, detail level, emotional tone, consistency, and imagery.
 \vspace{0.7em}
 
 \textbf{Stage 2: Human-like Reasoning} \\
 \hspace*{1em} For each major cue (at least 2–3): \\
 \hspace*{2em} • Identify the \textbf{Cue (signal noticed)} — e.g., simple language, excessive detail, balanced pros/cons, professional photos. \\
 \hspace*{2em} • State the \textbf{Perception (interpretation)} — e.g., “signals authenticity,” “triggers suspicion,” “feels promotional.” \\
 \hspace*{2em} • Write the \textbf{Explanation (reasoning)} — why this signal influenced the classification, using natural language and, if possible, example-style quotes (e.g., “This sounds too polished, like marketing copy, so it feels fake”). \\
 \hspace*{1em} If multiple cues conflict, explain how one outweighed the other in the final classification.
 \vspace{0.7em}
 
 \textbf{Rules} \\
 \hspace*{1em} 1. Always start with \textbf{Classification: Real/Fake}. \\
 \hspace*{1em} 2. Then provide at least 2–3 Cue $\rightarrow$ Perception $\rightarrow$ Explanation chains. \\
 \hspace*{1em} 3. Keep explanations natural and human-like, not abstract. \\
 \hspace*{1em} 4. If cues conflict, explain how one outweighed the other in the final classification.
 \end{tcolorbox}
 \caption{Meta-prompt template designed to generate human-centered, domain-sensitive explanations.}
 \label{fig:authenticity-prompt}
\end{figure}

This operationalization step responds directly to the observation that conventional explanation methods in fake review detection predominantly surface token-level relevance, which may reflect statistical correlations learned by the model but not the inferential strategies used by humans. Rather than translating word-level attributions into natural language, our approach starts from the users’ domain models themselves and treats these as first-class explanatory primitives. This shift reframes explanation from exposing model internals to supporting human sense-making.

Building on this thematic analysis, we examined how it could be operationalized in an LLM-based explanation agent. The "\textit{Cue $\rightarrow$ Perception $\rightarrow$ Explanation}'' mapping translates perceived deception cues (see Table~\ref{tab:thematic-summary}) into a reusable schema: Observed cues (e.g., plain language, staged photography) were systematically linked to user perceptions (trust vs. suspicion), which then motivate explanatory rationales. 

This mapping serves two functions. First, it provides structured knowledge for embedding into the agent, ensuring that explanations align with how users themselves make sense of reviews. Second, it enables the agent to produce transparent, human-like reasoning paths that mirror natural justificatory practices, rather than abstract, model-internal feature attributions. For example, instead of stating that a review was flagged due to lexical density, the agent could explain: \textit{"The language feels too polished and formal, which many people interpret as marketing rather than genuine feedback”}. 

Taken together, this process resulted in the development of a \textit{meta-prompt}, which operationalizes participants’ domain models into a structured instruction template. The final version of this meta-prompt is illustrated in Figure~\ref{fig:authenticity-prompt}.

\subsection{Adapting Explanations to Explanatory Objectives}
In the next step, we examined whether adapting the meta-prompt to different objectives changed how participants understood and used explanations. The structured reasoning path anchored outputs in empirically derived heuristics (linguistic style, detail, and reviewer credibility) while allowing objective-specific tailoring:
\begin{itemize}
 \item{\textbf{Understandability (Intelligibility)} {--} Explanations were concise and used plain language, echoing everyday heuristics (e.g., \textit{"This review uses casual wording, often linked to real experiences".})} 
 \item{\textbf{Trustworthiness (Accountability)} {--} Explanations highlighted motive attribution (e.g., \textit{"The repeated urging to buy suggests persuasive intent rather than authentic sharing".})} 
 \item{\textbf{Actionable (Persuasion-resistance)} {--} Explanations surfaced subtle manipulative cues and offered counter-arguments that participants could directly apply when judging reviews. } 
\end{itemize}

The juxtaposition of language-based explanations and word-highlighting explanations is not intended as a direct comparison of explanation formats in general, but as an examination of their respective alignment with users’ mental models in the fake review context. While word highlighting remains a widely used and valuable technique in text-based XAI, our study investigates whether such explanations provide meaningful support for credibility judgments when lexical overlap between classes is high. This comparison therefore serves to evaluate explanatory adequacy relative to human reasoning rather than technical fidelity alone.

\begin{figure}[!htb]
 \centering
 \begin{tcolorbox}[colback=gray!10, colframe=gray!40, width=0.95\textwidth, boxrule=0pt, arc=2pt, left=4pt, right=4pt]
 \small
 You are an XAI system \textbf{for a web shop (e.g., Amazon}). \\
 \\
 Based on the attached input, create a\textbf{convincing, trustworthy explanation in simple language} for a \textbf{web shop user who needs to quickly understand} why you classified the review as fake or genuine” \\
 
 \textit{<ATTACHMENT: OUTPUT OF THE DOMAIN EXPLANATION AGENT>}
 
 \end{tcolorbox}
 \caption{Meta-Prompt designed to generate context-specific explanations tailored to the intended target group.}
 \label{fig:user-facing-prompt}
\end{figure}

\begin{figure}[!htb]
\centering

\subfloat[Language-based explanation (User Explanation Agent).]{
    \includegraphics[width=0.8\linewidth]{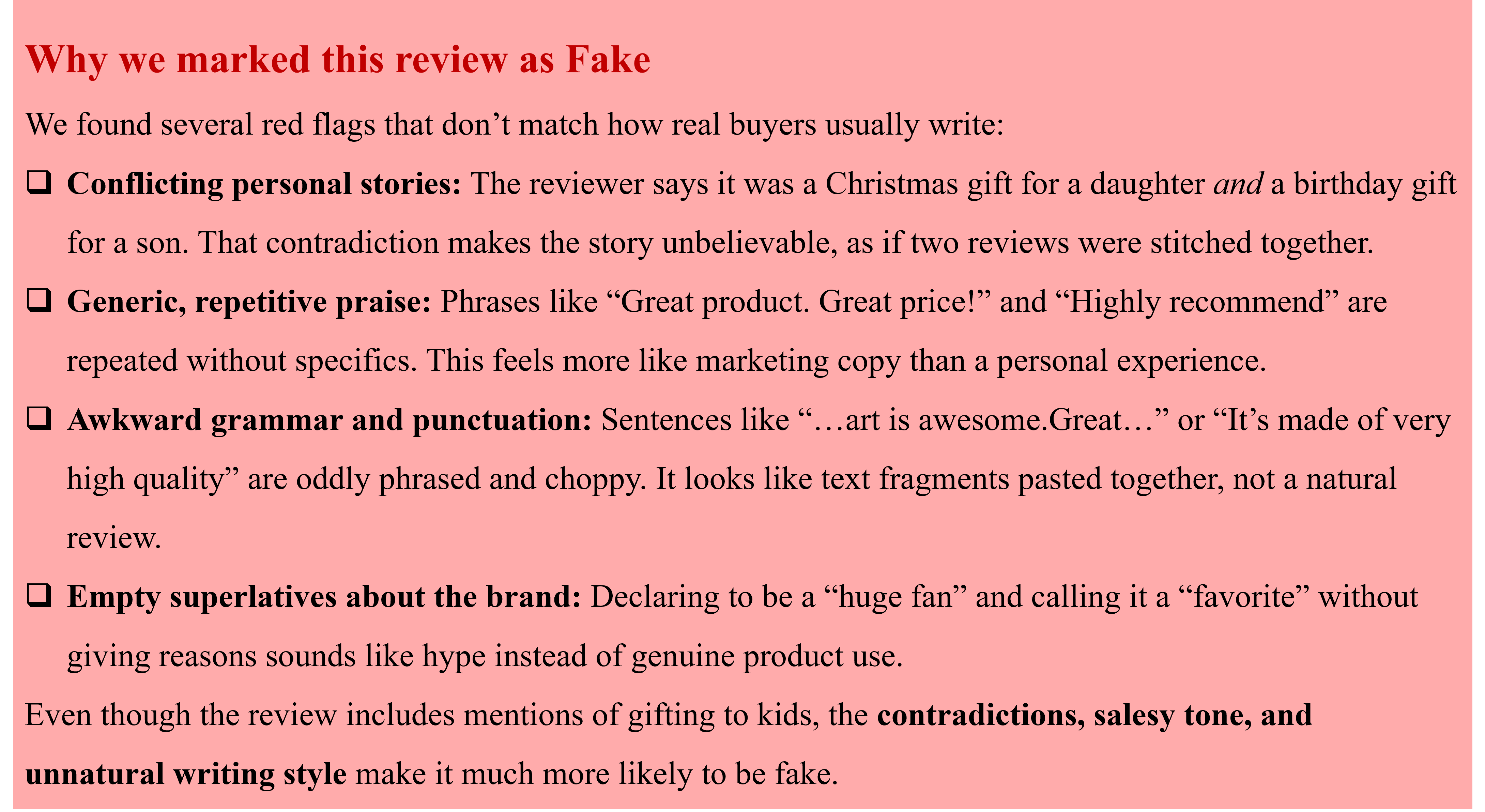}
    \label{fig:language}
}

\vspace{0.5cm}

\subfloat[Benchmark word-highlighting explanation.]{
    \includegraphics[width=0.8\linewidth]{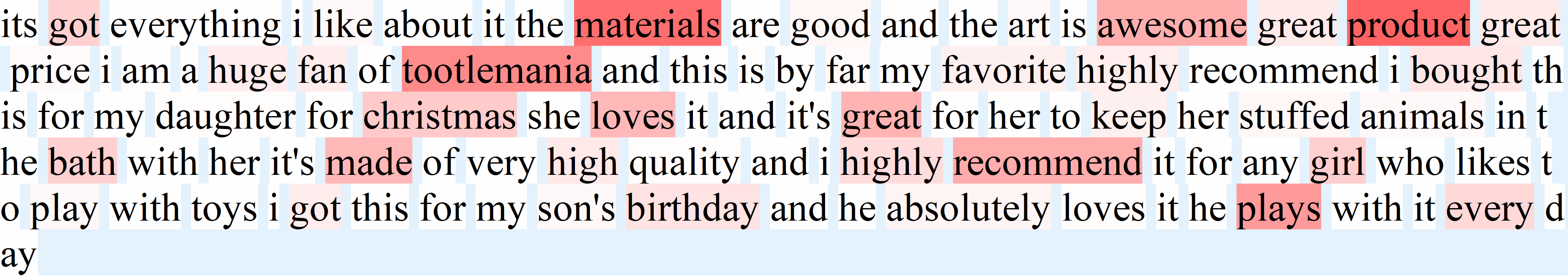}
    \label{fig:highlighting}
}

\caption{Two explanation types for fake review detection. Top: propositional, checklist-like rationales; Bottom: in-situ visual cues.}
\label{fig:interface}
\end{figure}

To operationalize this tailoring, we designed a secondary meta-prompt that recontextualizes the domain agent’s structured reasoning into
context-specific explanations for the intended target group. Iterative prompt engineering allowed us to probe the trade-off between fidelity and accessibility. The simple meta-prompt shown in
Figure~\ref{fig:user-facing-prompt} yielded the most effective outcomes, producing explanations that seems to be both cognitively lightweight for end-users and aligned with the mental domain models outlined above.

\section{Explanation Interface} \label{sec:heuristics}

It is important to emphasize that this comparison does not frame language-based explanations as universally superior to word-highlighting interfaces. Rather, it examines how different interface designs privilege different explanatory signals and cognitive processes. Word highlighting has proven effective in domains where class-related lexical features are discriminative and perceptually salient. In the fake review context, however, where authentic and deceptive reviews often share similar vocabulary, the adequacy of explanation interfaces must be assessed relative to users’ reasoning practices rather than their visual efficiency alone.

The User Explanation Agent generates a language-based explanation (Figure~\ref{fig:language}, top). In contrast, Figure~\ref{fig:highlighting} (bottom) shows the benchmark word-highlighting explanation used by Shajalal et al.~\citep{shajalal2024matters}. As \emph{explanation interfaces}~\citep{gunning2019darpa}, these variants make distinct design commitments (see Table~\ref{tab:design-comparison}).

\begin{table}[!htb]

\centering
\caption{Comparison structured by visual design, semantic--cognitive, and application levels.}
\label{tab:design-comparison}


\small
\begin{tabular}{>{\raggedright\arraybackslash}p{2.5cm} >{\raggedright\arraybackslash}p{5cm} >{\raggedright\arraybackslash}p{5cm}}
\toprule
\textbf{Issue} & \textbf{Language-based (top)} & \textbf{Word-highlighting (bottom)} \\
\midrule
\multicolumn{3}{l}{\textit{Visual design}}\\
& \textbullet\ Checklist-like bullet points
& \textbullet\ In-situ, token-level saliency overlay \\
& \textbullet\ Textual marking of key arguments
& \textbullet\ Graphical highlighting of salient words \\
& \textbullet\ Green/red-coded classification (entire box)
& \textbullet\ Green/red-coded classification (word-based) \\
\midrule
\multicolumn{3}{l}{\textit{Space design}}\\
& \textbullet\ Separate panel; detached from input
& \textbullet\ Co-located with the source text \\
& \textbullet\ Space-inefficient layout
& \textbullet\ Space-efficient layout; mobile-friendly \\
& \textbullet\ Requires spatial remapping
& \textbullet\ No spatial remapping \\
\midrule
\multicolumn{3}{l}{\textit{Semantic design}}\\
& \textbullet\ Propositional, symbolic
& \textbullet\ Indexical, deictic \\
& \textbullet\ Meaning conveyed via narrative scaffolding
& \textbullet\ Meaning conveyed via visual cues \\
& \textbullet\ Effortful, analytic processing \textit{(central route)}
& \textbullet\ Fast, heuristic processing \textit{(peripheral route)} \\
& \textbullet\ Supports symbolic analysis and abstract models
& \textbullet\ Supports visual analysis and preattentive ``pop-out'' \\
\midrule
\multicolumn{3}{l}{\textit{Application context}}\\
& \textbullet\ Propositional explanatory signals dominant (concepts/relations)
& \textbullet\ Perceptual explanatory signals dominant (visual patterns) \\
& \textbullet\ Justificatory, accountable; reflective tasks
& \textbullet\ Immediate, glanceable; visual tasks \\
\bottomrule
\end{tabular}
\end{table}

From a human-centered perspective, explanation interfaces are not neutral carriers of information but actively shape which aspects of a decision users attend to and how they interpret them. Consequently, interface design must be evaluated in relation to users’ mental models and the types of cues they consider meaningful. This framing aligns the interface comparison with our earlier qualitative findings, which showed that credibility judgments in fake review detection rely on higher-level conceptual cues rather than isolated lexical markers.

Regarding \textit{visual design}, both interfaces use green/red color-coding to encode classification information~\citep{bertin1983semiology}. The language-based design applies the color at the container (box) level, whereas the highlighting design applies it at the token (word) level within the text. With respect to visual structure, the language-based variant presents rationales in a bullet-point format, enabling users to quickly scan and verify claims much like a diagnostic checklist. The highlighting variant exploits pre-attentive processing~\citep{ware2012information} by embedding saliency directly in the text via a gradient scale. This supports rapid pattern recognition but yields only a coarse “less–more’’ spectrum. Through color saturation, lexical tokens are visually marked as points of importance, leaving users to infer their significance.

Regarding \textit{space design}, the language-based design separates the explanation from the source text, requires additional screen real estate, and introduces a spatial remapping between rationales and the review content. In contrast, the highlighting design is \emph{co-located} with the text and therefore more \emph{space-efficient} than a separate rationale panel—an advantage for constrained layouts (e.g., mobile).

Regarding \textit{semantic design}, the language-based explanation operates in a propositional mode, offering explicit rationales (e.g., “conflicting stories,’’ “awkward grammar’’). Such categorical statements frame the system as an expert narrator that articulates reasons for its judgment. By contrast, the highlighting modality relies on visual variables to signal potentially relevant parts of the text~\citep{bertin1983semiology}. Drawing on dual-process perspectives~\citep{petty1986elaboration}, the language-based explanation resembles the \emph{central route} of persuasion, encouraging users to integrate explicit rationales into analytic mental models. The highlighting explanation instead resembles a more \emph{peripheral} mode of processing, appealing to heuristic, perception-driven judgments by making certain tokens visually stand out. This distinction also resonates with XAI research: natural-language rationales can scaffold interpretability~\citep{miller2019explanation, zytek2024explingo}, whereas saliency-based explanations emphasize immediacy and transparency by visualizing signals associated with model decisions~\citep{lipton2018mythos}. The effectiveness of word-level highlighting further depends on whether nearby tokens can be perceived as meaningful units; otherwise, users must cognitively integrate dispersed signals across the text.

Crucially, this analysis does not suggest that word-highlighting explanations are inherently flawed. On the contrary, their strength lies in tasks where localized visual emphasis corresponds to meaningful explanatory units. However, in fake review detection, the absence of stable word-level discriminators means that highlighted tokens may appear arbitrary or misleading to end users. In such cases, interfaces that foreground conceptual relations and narrative rationales may better align with how users construct explanations and assess credibility.
\section{Evaluation Study}

Our design analysis indicates that the two explanation types operate on different semiotic and cognitive levels and thus exhibit context-specific strengths and limitations. Accordingly, the evaluation study investigated which explanation type is better suited for the domain of fake-review detection.
\\

\noindent\textbf{{\textit{Scope and Interpretation of the Evaluation.}}}
The goal of this evaluation was not to compare two explanation techniques as interchangeable variants of the same representational form, but to assess their suitability for the specific task ecology of fake review detection. Word- or token-level highlighting has been shown to be effective in domains where class-discriminative features correspond to stable, human-recognizable word sets (e.g., sentiment classification or topic attribution). In the fake review context, however, authentic and deceptive reviews often share the same vocabulary, while judgments hinge on higher-level semantic, stylistic, and pragmatic cues (e.g., tone, balance, coherence, persuasive intent). The comparison therefore evaluates whether explanations aligned with empirically elicited user mental models provide greater decision support than model-centric saliency cues in a domain where lexical attribution is structurally limited. Results should be interpreted with this task-specific framing in mind.

\subsection{Statistical Results}
\subsubsection{Construct and Discriminant Validity}
\label{sec:Validity}
To examine whether our measurement model was psychometrically sound, we conducted a confirmatory factor analysis (CFA) on the six explanation quality constructs adapted from prior XAI evaluation research~\citep{naveed2024overview}. Convergent validity was assessed via the average variance extracted (AVE), and internal consistency was examined using composite reliability (CR) and Cronbach’s~$\alpha$, following established benchmarks (AVE~$\geq .50$, CR~$\geq .70$, $\alpha \geq .70$)~\citep{hair2010multivariate}. Discriminant validity was evaluated using the Fornell--Larcker criterion~\citep{fornell1981evaluating} and the heterotrait--monotrait (HTMT) ratios~\citep{hair2010multivariate}. 

\begin{table}[!htb]
\centering
\caption{Convergent validity indices (sorted by AVE, then CR, then Cronbach’s $\alpha$).}
\label{tab:ave}


\begin{threeparttable}
\begin{tabular}{lccc}
\toprule
\textbf{Construct} & \textbf{AVE} & \textbf{Composite Reliability} & \textbf{Cronbach’s $\alpha$} \\
\midrule
Cognitive Load & .79 & .88 & .87 \\
Satisfaction & .75 & .86 & .86 \\
Decision Support & .74 & .85 & .85 \\
Trust & .71 & .83 & .83 \\
Understandability & .71 & .88 & .88 \\
Persuasiveness & .59 & .73 & .71 \\
\bottomrule
\end{tabular}
\end{threeparttable}
\end{table}

As shown in Table~\ref{tab:ave}, convergent validity was acceptable to good. All constructs met the recommended AVE~$\geq .50$ threshold, ranging from Persuasiveness ($.59$) to Cognitive Load ($.79$). Composite reliability values ($.73$–$.88$) and Cronbach’s~$\alpha$ coefficients ($.71$–$.88$) were also satisfactory, indicating internally consistent measurement scales~\citep{hair2010multivariate}. 

\begin{table}[!htb]
\centering
\small 
\caption{Discriminant validity indices}
\label{tab:discriminant}




\begin{threeparttable}
\begin{tabular}{lcccccc @{\hspace{1cm}} lcccccc}
\toprule
& Cl & Sa & Ds & Tr & Un & Pe & & Cl & Sa & Ds & Tr & Un & Pe \\
\midrule
Cl & \textbf{.89} & & & & & & Cl & -- & & & & & \\
Sa & .35 & \textbf{.87} & & & & & Sa & .36 & -- & & & & \\
Ds & .32 & .93 & \textbf{.86} & & & & Ds & .33 & .92 & -- & & & \\
Tr & .33 & .95 & .96 & \textbf{.84} & & & Tr & .34 & .95 & .96 & -- & & \\
Un & .33 & .99 & .95 & .97 & \textbf{.84} & & Un & .34 & .99 & .95 & .97 & -- & \\
Pe & .21 & .88 & .95 & .88 & .87 & \textbf{.76} & Pe & .21 & .89 & .96 & .89 & .88 & -- \\
\bottomrule
\end{tabular}
\begin{tablenotes}[para]
\footnotesize
\textit{Note.} Left panel: Fornell--Larcker criterion (diagonal $= \sqrt{\text{AVE}}$, lower triangle $=$ latent correlations $r$). Right panel: HTMT ratios (lower triangle). Abbr.: Cl: Cognitive Load; Sa: Satisfaction; Ds: Decision Support; Tr: Trust; Un: Understandability; Pe: Persuasiveness. 
\end{tablenotes}
\end{threeparttable}
\end{table}

Discriminant validity, however, was insufficient (Table~\ref{tab:discriminant}). Several constructs exhibited very high intercorrelations that exceeded the Fornell--Larcker threshold (e.g., Satisfaction and Understandability: $r\approx .99$ vs.\ $\sqrt{\text{AVE}}=.87$ and $.84$; Decision Support and Trust: $r\approx .96$). HTMT ratios further supported this lack of discriminant validity, with many values exceeding the recommended cutoff of $.85$–$.90$~\citep{henseler2015new}. 
This indicates that user appraisals such as Satisfaction, Trust, Understandability, and Decision Support overlapped strongly, consistent with the halo effects often observed in post-use evaluations~\citep{sauro2012quantifying}. Cognitive Load was comparatively distinct, showing only modest correlations with the other constructs ($|r|\approx .21$–$.35$). 

Despite the empirical overlap, we retained the theorized constructs in subsequent analyses. This decision is defensible for two reasons: (1) the constructs demonstrated acceptable convergent validity and reliability, and (2) prior work in XAI and HCI emphasizes their conceptual distinctiveness and practical relevance for explanation design. Still, high intercorrelations imply potential multicollinearity, so results should be interpreted with caution, and construct-specific differences should be regarded as provisional.

\subsubsection{Hypothesis Tests}
\begin{table}[!htb]
\centering
\small
\caption{Descriptive statistics and one-sample $t$-tests for explanation quality constructs (\(n_p=166\), \(n_r=996\)).}
\label{tab:apa-results}



\begin{threeparttable}
\begin{tabular}{
 @{} l
 S[table-format=1.2] 
 S[table-format=1.2] 
 c 
 S[table-format=2.2] 
 c c c 
 S[table-format=1.2] 
 @{}
}
\toprule
\textbf{Construct}
 & \multicolumn{1}{c}{\textbf{$M$}}
 & \multicolumn{1}{c}{\textbf{$SD$}}
 & \textbf{95\% CI}
 & \multicolumn{1}{c}{$t(995)$}
 & $p$ & $p_{\text{bonf}}$ & $p_{\text{FDR}}$
 & \multicolumn{1}{c}{$d$} \\
\midrule
Satisfaction & 1.33 & 1.04 & [1.26, 1.39] & 40.23 & $<.001$ & $<.001$ & $<.001$ & 1.27 \\
Understandability & 1.31 & 1.02 & [1.25, 1.38] & 40.76 & $<.001$ & $<.001$ & $<.001$ & 1.29 \\
Trust & 1.22 & 1.03 & [1.16, 1.29] & 37.56 & $<.001$ & $<.001$ & $<.001$ & 1.19 \\
Decision Support & 1.20 & 1.06 & [1.13, 1.26] & 35.70 & $<.001$ & $<.001$ & $<.001$ & 1.13 \\
Persuasiveness & 1.04 & 1.01 & [0.98, 1.10] & 32.55 & $<.001$ & $<.001$ & $<.001$ & 1.03 \\
Cognitive Load\tnote{a} & 0.83 & 1.25 & [0.75, 0.91] & 20.97 & $<.001$ & $<.001$ & $<.001$ & 0.66 \\
\bottomrule
\end{tabular}
\begin{tablenotes}[para]
\footnotesize
\textit{Note.} Responses were coded on a 5-point comparative scale and recoded to \(-2\) (``Much better Word-highlighting'') to \(+2\) (``Much better Language-based''); \(0\) indicates no difference. Positive means therefore favor language-based explanations. One-sample \(t\)-tests were conducted against the neutral midpoint \(0\). Effect sizes are Cohen’s \(d\) (one-sample). \(n_p\) = number of participants; \(n_r\) = total number of evaluated reviews (items aggregated across participants). Reverse coding was applied to \textit{Cognitive Load}, so that higher values indicate \emph{less} perceived mental effort for Language-based explanations compared to Word-highlighting.
\end{tablenotes}
\end{threeparttable}
\end{table}

The primary objective of this study was to evaluate the hypothesis that human-centered, language-based explanations generated by large language models (LLMs) would outperform a word-highlighting baseline across six explanation-quality dimensions. One-sample $t$-tests were conducted against the neutral midpoint ($0=\text{``about the same''}$), testing the directional hypothesis that language-based explanations would be rated more favorably. To account for multiple comparisons across the six constructs derived \textit{a priori} from the literature, Bonferroni corrections were applied ($\alpha_{\text{adj}}=0.05/6\approx0.008$). False-discovery-rate (FDR)--adjusted $p$-values were also computed for robustness, yielding consistent results: all effects remained statistically significant. Results, summarized in Table~\ref{tab:apa-results}, confirmed all hypotheses, with language-based explanations rated significantly higher across all dimensions. Effect sizes are reported as Cohen's $d$ with 95\% confidence intervals for the mean. By Cohen’s conventional benchmarks for \(d\) (0.2 small, 0.5 medium, 0.8 large; \citep{hair2010multivariate}), the effects were predominantly large: \(d\) ranged from 0.66 to 1.29 (see Table~\ref{tab:apa-results}), with all constructs except Cognitive Load (\(d=0.66\)) exceeding 0.80.

Specifically, language-based explanations were most favored for Satisfaction ($M=1.33$, $d=1.27$) and Understandability ($M=1.31$, $d=1.29$), indicating substantial improvements in perceived experiential quality. Trust ($M=1.22$, $d=1.19$) and Decision Support ($M=1.20$, $d=1.13$) also showed strong preferences, while Persuasiveness ($M=1.04$, $d=1.03$) and Cognitive Load ($M=0.83$, $d=0.66$, reverse-coded) exhibited somewhat smaller but still significant effects.
These findings provide robust support for the superiority of language-based explanations over word-highlighting across all evaluated dimensions. The rank order of constructs suggests that experiential qualities (satisfaction and understandability) were most enhanced, while cognitive load, though improved, showed the least pronounced effect relative to the baseline. However, due to poor discriminant validity, claims about the ranking remain provisional. To gain a deeper understanding of this trade-off and its implications, we analyze in the following section the qualitative feedback in detail.

\subsection{Qualitative Feedback}
\subsubsection{General Feedback} 

The qualitative analysis of the open-ended questions supports and extends the statistical findings of the A/B test by offering insight into how participants experienced and evaluated the two explanation formats. 

Across responses, participants consistently emphasized the advantages of the language-based explanations, noting their clarity, comprehensibility, and transparency of reasoning, while also identifying potential areas for refinement. When explicitly asked about comparative advantages, participants overwhelmingly reported that the language-based approach was superior, whereas only a small minority attributed any benefits to word-highlighting, usually in terms of speed or efficiency rather than explanatory adequacy. 

Participants’ overall impressions strongly favored the language-based format, which they described as structured, coherent, and human-like in its provision of rationales rather than fragmented signals. As one participant expressed: \emph{``it provided a rational and in-depth explanation for the decision made, it explained the type of words/grammar used, any contradictions, and what would be expected from a real vs fake review''} (P704). By contrast, word-highlighting was frequently dismissed as opaque, confusing, or seemingly arbitrary. As another participant noted: \emph{``I found the word highlighting quite difficult to follow''} (P710). These contrasting accounts illustrate how users anchor their judgments of explanation quality not only in the communicative form but also in the perceived cognitive utility of the explanation, thereby underscoring the centrality of interpretability and transparency in shaping user trust and engagement.

\subsubsection{Feedback Themes} 
The thematic analysis of participants’ accounts highlights how explanation formats shape processes of sense-making, trust calibration, and user engagement. In this regard, we uncovered the following themes in the feedback:\\

\noindent\textbf{\textit{Clarity and Ease of Understanding.}}
Clarity emerged as a decisive theme. Participants repeatedly emphasized that the narrative style of language-based explanations was easy to follow, written in accessible terms, and thus reduced the interpretive burden. As one participant summarized: \emph{``it is easy to read and to understand''} (P705). From a human–computer interaction perspective, this reflects the alignment of explanatory discourse with users’ natural interpretive strategies. In contrast, word-highlighting often imposed additional interpretive work, effectively shifting cognitive effort from the system to the user. This demonstrates a key design tension: visual simplicity does not necessarily equate to cognitive simplicity, as the latter requires reducing ambiguity and providing coherent interpretive scaffolds. \\

\noindent\textbf{\textit{Depth and Informational Richness.}} 
Beyond clarity, participants valued the informational richness of the language-based format. Explanations were perceived as embedding the system’s assessment within contextual reasoning, thus enabling participants to appropriate the rationale for their own judgments. The perceived epistemic depth illustrates the importance of explanations that do more than present surface features; they must situate those features in relation to broader linguistic or stylistic patterns. Word-highlighting, by contrast, was experienced as shallow—pointing to lexical elements without articulating why they mattered. This underscores the distinction between feature-level transparency and genuine interpretability: visibility of features alone does not constitute an explanation in the sense required for effective user sense-making. \\

\noindent\textbf{\textit{Transparency and Logic.}} 
Participants often described the narrative explanations as transparent and logically coherent, in contrast to the perceived arbitrariness of highlighting. One participant articulated this contrast explicitly: \emph{``Language based is straightforward, human-like and easy to digest while word highlighting ... I cannot understand what it means and what is a reason between all that highlight''} (P703). This observation resonates with HCI research emphasizing that transparency depends not simply on exposing system internals, but on communicating reasoning in a way that is intelligible and accountable to users. \\

\noindent\textbf{\textit{Trust and Persuasiveness.}} 
The persuasive quality of language-based explanations was another salient theme. Participants reported that the detail and coherence of these accounts enhanced their trust in the system’s decision. Explanations were not only clearer but also more convincing, producing what might be termed “persuasive transparency.” Conversely, the arbitrariness of highlighting undermined credibility and raised doubts about the classification process. These findings align with HCI literature that views explanations as central to trust calibration: they are not neutral artifacts but active mediators of user confidence. \\

\noindent\textbf{\textit{Criticism of Word-highlighting.}}
Word-highlighting was frequently criticized as an underdeveloped explanation strategy. Participants described it as incomplete, unhelpful, or even misleading. The absence of causal or logical connections between highlighted features and classification outcomes was particularly problematic. In HCI terms, highlighting offered a “what” without a “why,” leaving participants without a reconstructable reasoning process. This critique underscores that explanation strategies must go beyond surface-level visibility to support interpretive accountability.\\ 

\noindent\textbf{\textit{Minor Positive Mentions of Word-highlighting.}} 
Although the vast majority of participants evaluated word-highlighting negatively, a small number acknowledged its potential efficiency benefits. In these accounts, highlighting was valued as a lightweight heuristic—useful for rapid consumption when depth was not required. Such perspectives suggest that while word-highlighting is insufficient as a stand-alone strategy, it may contribute as part of a hybrid explanation design that flexibly accommodates diverse user needs and contexts. \\

\noindent\textbf{\textit{Explanatory Anchors.}} 
A prominent theme was that participants oriented to explanations as simultaneously anchored in the AI’s reasoning and in the review text as the object of that reasoning. Rather than treating these orientations as distinct, participants often tied the anchors together. They perceived the explanation both as an account of the features of the (fake) review and as an account of the features considered by the AI in making its judgment
Some participants foregrounded the review as the primary object of evaluation, treating the explanation as a tool for judging its textual qualities. One participant, for example, remarked: \emph{``it shows whether the review is good''} (P711). Others emphasized the reasoning of the AI system, focusing on how explanations could clarify the decision process itself. As one noted, \emph{``I would like a short note next to the highlights saying why they are relevant''} (P704). Many responses, however, blended these orientations by linking linguistic or stylistic markers in the review to the classifier’s evaluative logic. One participant captured this integration clearly: \emph{``it provided a rational and in-depth explanation for the decision that was made, explaining the type of words/grammar used, any contradictions, and what would be expected from a real vs fake review''} (P704).

\subsubsection{Areas for Improvement} 
We invited participants to provide open-ended feedback on areas for improvement in both explanation types. The following section summarizes the main insights.\\

\noindent\textbf{\textit{Improving the Language-based explanations.}} 
While the narrative format was strongly preferred, participants identified areas for refinement. Some described the texts as too verbose or repetitive, noting that excessive detail risked producing cognitive overload. Suggestions included shorter, more concise narratives that foreground key points, as well as clearer structuring through emphasis or formatting. Several participants also proposed hybrid approaches, combining narrative explanations with selective highlighting to balance richness of reasoning with efficiency of consumption. These reflections illustrate a core challenge in explanation design: reconciling depth and detail with accessibility and usability. \\

\noindent\textbf{\textit{Improving the Word-highlighting explanations.}} 
Participants’ recommendations for improving word-highlighting were more fundamental. The dominant critique was that highlighting failed to provide explanatory value without explicit connections to the underlying reasoning. Participants suggested making highlighting more selective and targeted, reducing visual noise, and embedding annotations to contextualize why specific words mattered. These proposals underscore that highlighting cannot be treated as a self-sufficient explanation but requires integration with explanatory narratives to become meaningful.

\section{Limitations}
\label{sec:limitation}

In this paper, we demonstrate that our general framework, \emph{MIND-X}, for human-centered explanation design outperforms existing approaches. However, the empirical evidence is derived from a single application scenario, and its transferability to other domains, modalities, and languages therefore remains an open question. In addition, our elicitation of users’ mental models relies on a small and relatively homogeneous sample of highly educated participants, which limits representativeness. While small samples are common in qualitative research, and our quantitative evaluation suggests that the resulting taxonomy provides a sufficient basis for generating high-quality explanations, broader and more diverse samples would strengthen the robustness of these findings.

In the quantitative evaluation, we adopt perceived qualities as evaluation metrics established in the literature~\citep{hoffman2018metrics, naveed2024overview}. However, behavioral outcomes and longitudinal effects (e.g., changes in actual shopping behavior) fall outside the scope of this study. Another limitation concerns psychometric considerations. Although rigorous construct validation remains relatively rare in XAI studies, we followed APA guidelines and conducted measurement analyses. These analyses revealed weak discriminant validity and substantial inter-correlations among some scales. While this does not affect our central finding, such as users’ stronger preference for the proposed approach, it highlights the need for more robust and orthogonal measurement instruments in future work.

Because our evaluation focuses on perceived qualities, we do not assess formal properties of the explanations with respect to the underlying AI model, such as faithfulness or robustness. While such metrics are central in technical XAI research~\citep{doshi2017towards}, a human-centered perspective emphasizes how users perceive explanations and make sense of them. At the same time, this limitation points to a broader disciplinary gap and highlights the ongoing tension between faithfulness and usefulness in explanation design.

\section{Discussion}

\subsection{Beyond Model-Centric Explanations: Tailoring to Users’ Mental Models}
Explanation approaches in XAI \citep{doshi2017towards} have traditionally relied on feature-attribution techniques \citep{ribeiro_lime_2016} and example-based methods such as counterfactuals \citep{dwivedi2023explainable} to elucidate how inputs influence model decisions. More recent work has shifted toward concept-based explanations \citep{kim2018interpretability} and LLM-generated natural-language rationales, aiming to improve comprehensibility. 

Building on this trajectory, our framework integrates conceptual and narrative explanations \citep{slack2023explaining} with a human-centered design stance \citep{liao2020questioning, chromik2021human}. Moreover, \emph{MIND-X} extends these contributions by providing a six-step, reproducible pipeline that operationalizes them end-to-end, showing how explanations can be aligned with a \emph{user domain model} of the concepts lay users actually reason with. This represents a shift from \textit{model-centric transparency} to human-centered \emph{account-giving} grounded in users’ domain mental models.

A core aim of \emph{MIND-X} is to move beyond programmatic claims by 
demonstrating, through a worked example, how human-centered design practices can be systematically integrated with contemporary XAI techniques. Concretely, we demonstrate how common HCI methods \citep{norman2013design} can be adopted to elicit and formalize users’ domain heuristics, and how this knowledge can be used to design structured meta-prompts and explanation schemas \citep{qiu2024scenarios, slack2023explaining}. Regarding explanation design, a central architectural contribution of \emph{MIND-X} is the explicit separation of \emph{what to explain} from \emph{how to explain}. By aligning content selection with user- and domain-facing concepts before choosing the communicative form (e.g., narrative rationales, in-situ highlights), the aim of \emph{MIND-X} is to ensure semantic grounding before determining how explanations are presented.This advances needs-first guidance that surfaces explanatory goals but leaves their operationalization under-specified \citep{abdul2018trends}, and extends design-principle syntheses that frame explanations as interactive resources \citep{chromik2021human, eiband2018transparency} by providing a procedural pipeline with explicit tailoring levers (tone, granularity, framing). It further complements concept-based interpretability that centers model-internal concepts \citep{kim2018interpretability, koh2020concept} and LLM-generated rationales \citep{slack2023explaining, zytek2024explingo} by binding concept selection to user-domain models and evaluating both model-consistency and domain-justification.

Empirically, \emph{MIND-X} generated explanations showed advantages over token-level word highlighting \citep{kim-etal-2021-distinguish,shajalal2025dothe}. In our benchmark evaluation, we assess six established metrics (understandability, satisfaction, trust, decision support, persuasiveness, perceived cognitive load) commonly used in human-grounded evaluation \citep{hoffman2018metrics}. Across all six outcomes, participants showed a strong preference for language-based explanations, and all effects remained significant after controlling for multiple comparisons \citep{hair2010multivariate}. The largest gains appeared for satisfaction, understandability, and trust; the smallest (yet still significant) improvement concerned perceived cognitive load. Because discriminant validity among scales is weak, we interpret the \emph{rank ordering} of constructs cautiously due to potential halo effects \citep{henseler2015new}, while our overall superiority claim remains unaffected (each dimension independently favored language-based explanations). This quantitative pattern aligns with our heuristic interface analysis (Section~\ref{sec:heuristics}) and with our qualitative evidence: participants emphasize clarity, traceability, and trust-building for language explanations, whereas highlighting was often described as “arbitrary” or explanation-poor. This resonates with Joshi et al.~\citep{joshi2023machine}, underscoring that explainability \emph{per se} does not guarantee \emph{actionable advantages} for lay users.

However, the strong user preference for our explanation approach should not be interpreted as a dismissal of token-level highlighting. On the contrary, our heuristic analysis (Section~\ref{sec:heuristics}) indicates that the two explanation types pursue different design goals and excel in different usage contexts. This view is supported by qualitative feedback from the evaluation study: several participants explicitly request \emph{hybrid} solutions, getting concise narratives complemented by targeted highlights. Taken together, these observations support a multi-modal design stance that combines propositional, narrative rationales \citep{zytek2024explingo} with deictic, visual cues \citep{bertin1983semiology}. This perspective aligns with current HCI/XAI work that treats explanations not as static representations but as \emph{interactive resources} supporting users’ ongoing sense-making \citep{chromik2021human, eiband2018transparency}.\\

\textbullet{\textit{Why This Comparison Is Task-Specific Rather Than Method-Generic?}}
A potential concern is that comparing narrative, language-based explanations with token-level highlighting constitutes an unfair or mismatched baseline. We emphasize that this comparison is intentional and task-specific rather than method-generic. Word-highlighting explanations are well-suited to domains where class membership hinges on stable, discriminative lexical features (e.g., sentiment analysis or topical classification). In fake review detection, by contrast, deceptive and genuine reviews frequently share vocabulary, and judgments depend on higher-level semantic, stylistic, and pragmatic cues. The comparison therefore probes explanation–task alignment: whether explanations grounded in user mental models better support sense-making in a domain where word-level attribution is structurally limited. We do not claim universal superiority of narrative explanations, but domain-appropriate adequacy.

Our findings highlight the need for human-centered, domain-specific explanation design, which produces \textit{necessarily incomplete}~\citep{keil2006explanation}, yet contextually sufficient accounts \textit{for all practical purposes}~\citep{garfinkel1967studies}. Supporting context-specific tailor-ability, the proposed approach extends existing approaches in a structured manner. It implements a concept-first separation of \emph{what to explain} (user-aligned concepts) from \emph{how to explain} (tailored narratives), echoing a concept-bottleneck \citep{koh2020concept} and needs-first rationale \citep{liao2021human}, while leveraging LLMs for adaptive outputs. 
This separation supports LLM-based explanation design~\citep{zytek2024explingo} by ensuring that explanation content is grounded before being adapted to different communicative contexts~\citep{chromik2021human}. As a result, explanations can remain semantically meaningful while also being recognizable and actionable in specific use contexts.

\subsection{Dealing with the Explanation Dilemma: Faithfulness vs.\ Situated Intelligibility}

The qualitative feedback also reveals that participants repeatedly characterize the narratives as structured, coherent, and human-like, while describing the highlighting baseline as opaque and seemingly arbitrary.
Placed in a wider epistemic frame, this observation directly challenges the positivist, causal-realist orientation that still underwrites much XAI work, where explanatory legitimacy is cast in correspondence terms: an explanation is “good” insofar as it \emph{faithfully} reflects internal model causality~\citep{doshi2017towards}. By contrast, social-constructivist accounts ask us to evaluate explanations with respect to their \emph{interpretant}—the meaning taken up by a human interpreter and their role in situated sense-making~\citep{weick1995sensemaking}. From this standpoint, explanations are not neutral mirrors but \emph{accounts} rendered intelligible and actionably adequate within specific practices.
From the causal-realist stance, however, we encounter a stark dilemma, well recognized in CSCW when formal systems meet lived practice \citep{suchman1987plans}:

\begin{itemize}
\item \textit{Horn A: Formally Faithful Yet Socially Unintelligible.} On the one hand, we have models and post hoc methods that boast formally demonstrated accuracy or faithfulness to their internal mechanisms (as our benchmark model does~\citep{shajalal2025dothe}). Yet, as our study demonstrates, these explanations risk being \emph{experienced} by users as opaque, arbitrary, or merely gestural, a \textit{transparency gloss} that reveals internals without rendering them accountable to the task at hand. The correspondence claim holds, but the account fails: the explanation does not travel from model semantics to human sense-making.
\item \textit{Horn B: Socially Intelligible Yet Epistemically Unsecured.} On the other hand, we have explanations that are reported as useful, persuasive, and intelligible—precisely because they are shaped by human explanatory practices~\citep{lipton2018mythos,miller2019explanation}. Yet their epistemic status is undetermined: their generative mechanisms are only partially understood, their quality is not guaranteed by formal criteria, and they are vulnerable to well-documented failure modes such as hallucination \citep{kunz2024properties}. Here the account succeeds in practice, but its warrant relative to the model’s actual inferential grounds remains uncertain.
\end{itemize}

This is not a superficial trade-off but an \emph{epistemic double bind}. The two horns index \emph{incommensurate accountability targets}: (i) accountability to the \emph{system’s} causal story (model-internal semantics) and (ii) accountability to the \emph{world of work} in which people must judge, justify, and act (domain-level sense-making). In CSCW terms, the former privileges formal representations; the latter privileges situated action and ethnomethodological accountability~\citep{schwartz2013uncovering, garfinkel1967studies}. Explanations confined to one pole predictably underperform at the other: model-faithful rationales that no one can use, or human-friendly stories that are weakly tethered to what the model in fact did. As Kunz and Kuhlmann \citep{kunz2024properties} caution, the absence of model-centric faithfulness can be dangerous if it results in persuasive but misleading accounts. Yet the inverse holds as well: an exclusive focus on faithfulness can itself be dangerous when it produces only a \emph{transparency gloss}, inviting a false sense of understanding while failing to support situated sense-making.

\subsection{From Correspondence to \emph{Double Accountability}: Bridging Model Reasoning and Domain Sense-Making} Our empirical results suggest a way through. Our study shows that explanations are judged neither as mere reflections of model logic nor as mere summaries of textual evidence, but as \emph{interpretive resources that must bridge both}. Moreover, our findings suggest that people do not treat explanations as either mirrors of model computation or mirrors of textual reality. Rather, they hold explanations to a \emph{dual standard of accountability}: simultaneously to the AI’s decision procedure (\textit{“Why does the classifier decide X?”}) and to the review as a world-facing object of judgment (\textit{“Why is this review fake/real?”}). Participants’ comparative evaluations and open-ended comments make clear that these orientations are not experienced as separable tasks but as a single, doubly oriented sense-making problem: a good explanation has to \emph{bridge} the two, showing how the review’s features could be read as supporting the system’s assessment and how the system’s assessment made the review intelligible as fake or genuine. 

Instead of relying on a single criterion of \textit{faithfulness}, a broader notion of explanatory adequacy is needed \citep{slack2023explaining}. In hybrid human–AI decision environments \citep{raikov2024accelerating}, explanations must support both the interpretation of model outputs and the interpretation of domain evidence.We therefore propose a \emph{two-dimensional conceptualization of explanatory adequacy}:
\emph{model-fit} (does the account render the system’s decision procedure intelligible and checkable?) and \emph{world-fit} (does the account embed the decision in recognizable, assessable domain evidence). 

In semiotic terms, adequacy is a property of the explanation–interpreter relation (its \emph{interpretant}), not solely of the explanation–model correspondence. In ethnomethodological terms, adequacy is achieved when an explanation functions as a \emph{work practice} for making a decision \emph{accountable} \citep{schwartz2013uncovering, garfinkel1967studies}, both to the classifier’s reasoning and to the object of judgment.

\subsection{Implications for Human-Centered XAI}
Our study allows us to make several suggestions for designing human-centered XAI systems. First, evaluation criteria must be re-framed. Metrics that certify \emph{only} internal correspondence undervalue the very qualities users report needing (sense-making, justificatory traction, actionability), while metrics that certify \emph{only} human preference risk licensing fluent but unfounded stories. A double-accountability regime therefore calls for paired assessments: stress-tests of model-consistency \emph{and} task-grounded tests of justificatory adequacy (can users reconstruct why \emph{this} evidence warrants \emph{this} decision, and contest it when it does not?).

Second, design should target \emph{account-giving} rather than mere \emph{feature-showing}. Narrative, cue-linked rationales can make the mapping from domain evidence to model judgment explicit, while selective highlighting or structured artifacts (e.g., evidence tables, contrastive cases) provide checkable anchors. In this hybrid, symbolic accounts perform the justificatory work; indexical cues keep the account tethered to inspectable evidence.
Finally, double accountability clarifies additional responsibilities: System developers owe users model-grounded warrants (avoiding unfalsifiable glosses); interface designers owe users situated intelligibility (avoiding alluring but unmoored stories); and evaluators owe the field protocols that test both anchors together.

\section{Conclusion and Outlook}
This paper advances explainable AI by introducing \emph{MIND-X}, a human-centered framework that grounds explanations in users’ mental models and common-sense reasoning rather than in model-centric attributions. 
Methodologically, MIND-X proceduralizes prior human-centered guidance into a \emph{six-step, reproducible pipeline} and treats LLMs as orchestrated \emph{explanation agents}, making design assumptions explicit, auditable, and portable across domains. 

Empirically, the case study in fake-review detection shows that narrative, heuristic-aligned accounts foster intelligibility, trust, and decision support, highlighting the limits of token-level transparency. Qualitative feedback further corroborates these results, revealing that users experienced narrative explanations as more insightful, both in terms of why reviews might be fake and why they were ultimately classified as such. In this regard, our study demonstrates the need for \textit{double accountability}—to both model logic and human sense-making.

Importantly, these findings should not be read as a general indictment of word-level attribution techniques. Rather, they underscore that explanation effectiveness is contingent on task structure and the nature of the evidence users rely on. While token-level highlighting may be effective in domains with stable, class-discriminative lexical cues, fake review detection relies heavily on higher-order semantic, stylistic, and pragmatic judgments. Our contribution therefore lies in demonstrating the importance of explanation–task alignment, rather than asserting a universally superior explanation modality.

By framing explanations as discursive accounts rather than technical artifacts, our aim is to move beyond the mere \textit{gloss of transparency} toward explanations that resonate with everyday reasoning and support actionable judgment.

Looking ahead, three directions for future research emerge. First, the emergence of large language models (LLMs) as agents of explanation raises concerns about consumer manipulation and the potential for ethics-washing through superficial explanatory glosses, thereby necessitating safeguards against persuasive yet weakly substantiated accounts. Second, although the study results show a strong preference for our approach, user feedback also indicates that design refinements should explore hybrid formats combining narrative rationales with selective visual cues. For this reason, future research should consider adaptive interfaces that dynamically tailor explanations to users’ expertise, goals, and situational constraints. Third, while our study focuses exclusively on subjective perception, taking the risk of misleading explanations seriously requires evaluating them not only through model-centric metrics or self-reported perceptions, but also in ecologically valid contexts that examine behavioral and longitudinal outcomes such as decision accuracy, resilience to manipulation, and long-term trust calibration.

In this way, future research can move toward a more comprehensive paradigm of explainable AI, one that reconciles technical fidelity with human intelligibility and social responsibility, upholds epistemic responsibility, equipping users with explanations that are not only transparent but also actionable in everyday decision-making.

\section*{Ethical Statement}
The study was conducted in accordance with the ethical standards of the University of Siegen and the principles of the 1964 Helsinki Declaration and its subsequent amendments. The Ethics Council for Research of the University of Siegen reviewed the project on 18 June 2026 and determined that no formal ethical approval was required, classifying the project as unobjectionable (Reference No.~LS\_ER\_34\_2026).

\section*{Acknowledgment}
We acknowledge the use of LLM for research assistance in this work. The LLM was utilized for editing, structuring text, improving language, and checking the logical consistency of the text. All content was reviewed, edited, and validated by the authors to ensure accuracy, originality, and compliance with the Taylor and Francis on Use of Large Language Models. The authors take full responsibility for the content and integrity of this work.

\section*{Funding}
This research is a part of COPP: Co-pilot in Production Project, funded by EFRE/JTF-Programm NRW, Germany, Grant reference EFRE-20801751.

\section*{Disclosure statement}
No potential conflict of interest was reported by the author(s).

\section*{Data Availability Statement}

The anonymized quantitative dataset supporting the findings of this study is openly available in Zenodo under a CC BY 4.0 licence at \url{https://doi.org/10.5281/zenodo.21185769}~\cite{shajalal2025mindxdata}. The repository contains the quantitative data underlying the statistical analyses reported in this article.

\section*{Author Contribution}
\textbf{Sidra Naveed:} Conceptualization, Methodology, Validation, Formal analysis, Investigation, Data curation, Writing—original draft preparation, Writing—review and editing, Visualization, Project administration; \textbf{Md Shajalal:} Conceptualization, Methodology, Software, Formal analysis, Investigation, Data curation, Writing—original draft preparation, Writing—review and editing, Visualization; \textbf{Gunnar Stevens:} Conceptualization, Methodology, Formal analysis, Investigation, Writing—original draft preparation, Writing—review and editing, Supervision, Project administration; \textbf{Alexander Boden:} Validation, Writing—review and editing, Supervision

\bibliographystyle{ACM-Reference-Format}
\bibliography{cleaned_references}

\appendix
\section{Reviews and Generated Explanations}\label{appendix}

\subsection{Review 1}
\noindent \textbf{Scenario: Sports and Outdoor Items (Gloves)}: You’ve just joined a local gym and need reliable workout gloves to prevent calluses and slipping during weight training. Before making a purchase, you’re browsing customer reviews to see if these gloves will hold up for regular exercise. 

\noindent \textbf{Customer Review}: 
``I actually have a review here on the site about these gloves. These gloves are awesome and I highly recommend them. I am a 32 year old man who works at a large company and have had no issues with these gloves. I have had no issues with the gloves falling apart or breaking. They are the most comfortable I have ever worn and I am so happy with them. I recommend them to anyone looking for a glove that will last a long time. I am very happy with my purchase and would recommend these gloves to anyone looking for a great glove. These are super lightweight and lightweight. I have used them for a couple of months and they are still working great. I would definitely recommend them to anyone looking for a good quality product. I've had no problems with them slipping or sliding. My only complaint is that I haven't used them yet, but they are very well made and have been used in my gym.''

\begin{figure}[!htb]
    \centering

    \subfloat[\textbf{Language based Explanation}\newline\newline\label{fig:R1Lang}]{
        \includegraphics[width=0.7\linewidth]{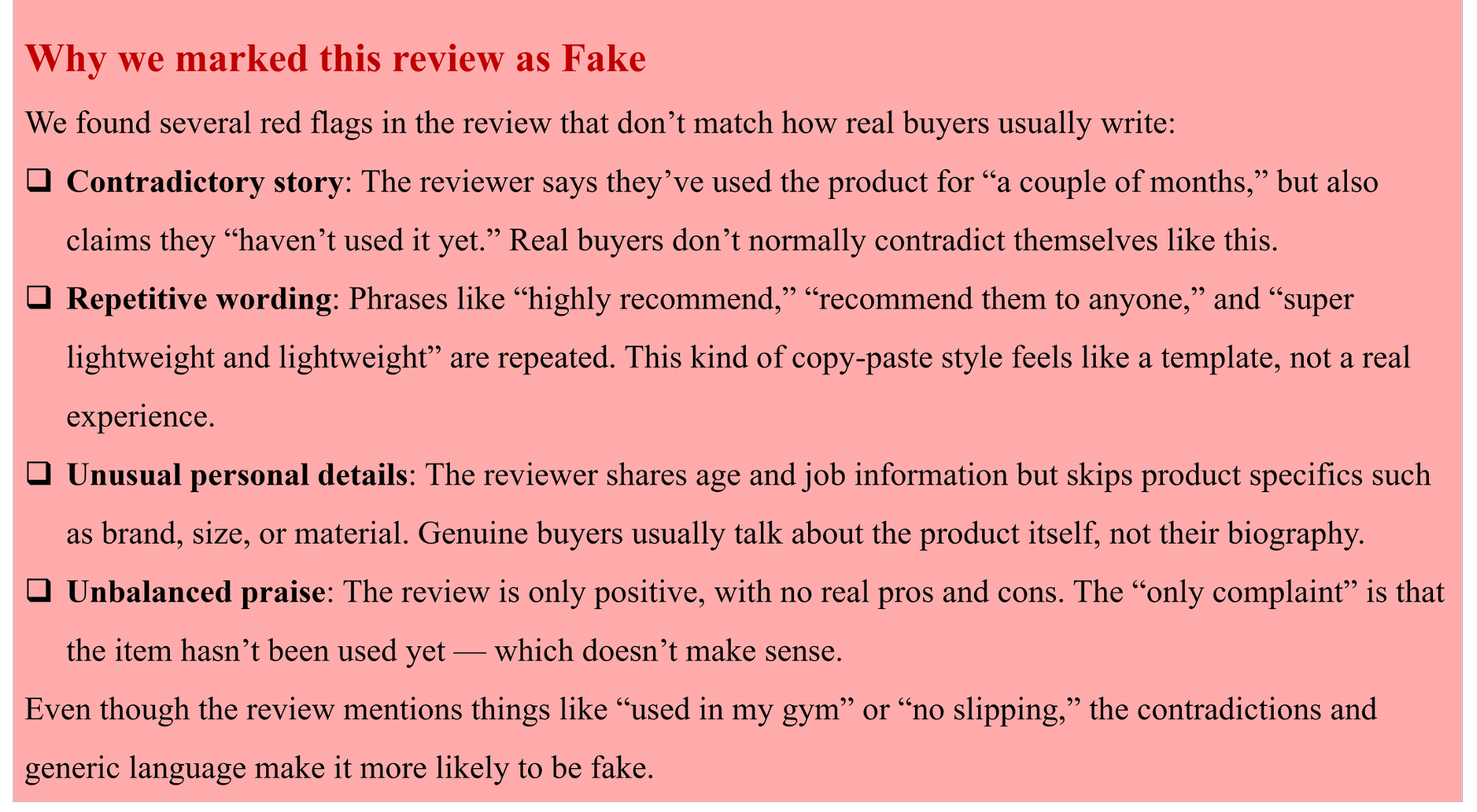}
    }

    \subfloat[\textbf{Word-highlighting based explanation}\label{fig:R1word}]{
        \includegraphics[width=0.7\linewidth]{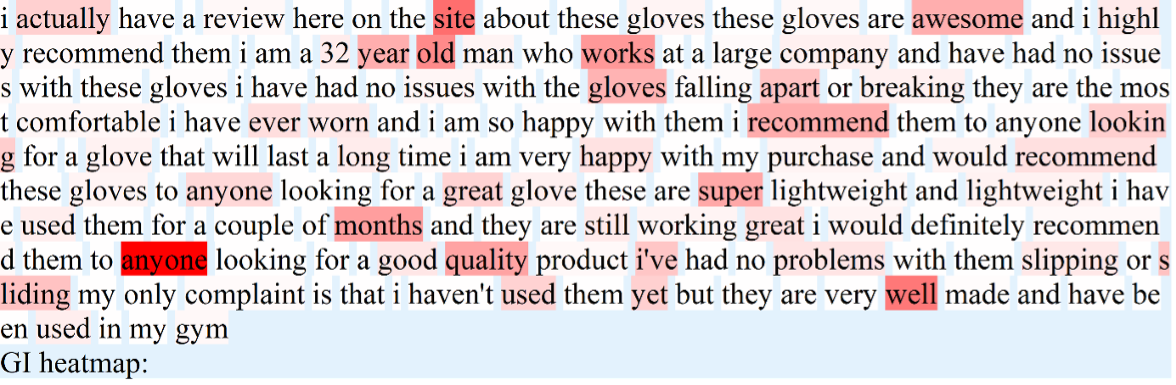}
    }

    \caption{Explanations generated by both systems for Review 1}
    \label{fig:R1Explanations}
\end{figure}
\newpage
\subsection{Review 2}
\noindent\textbf{Scenario: Pet Supplies (Cat Tree)}: You recently adopted a mother cat and her kittens. You’re looking for a sturdy and safe cat tree where they can climb and play. You decide to read reviews to understand whether this product will be suitable for playful kittens and a nursing cat.\\

\noindent\textbf{Customer Review:} 
``For a mother cat and three small kittens. Solid, easy to put together and pleasant to look at. Has many nooks and crannies and is well-spaced for the little ones. It's well balanced so they can jump around and I've seen no rocking or instability. The only downsides are that the insides of the two houses are not lined with fabric but just bare wood, which is not obvious in the photos or description, and the fabric itself (other than the sisal legs) is plush, not typical carpet. It's very slippery and they have a hard time getting there claws in it. I'd steer clear of this type of thing in the future and will go with normal carpeted types. Too bad, otherwise it's something I'd buy again.''\\

\begin{figure}[!htb]
    \centering

    \subfloat[\textbf{Language based Explanation}\newline\newline\label{fig:R2Lang}]{
        \includegraphics[width=0.7\linewidth]{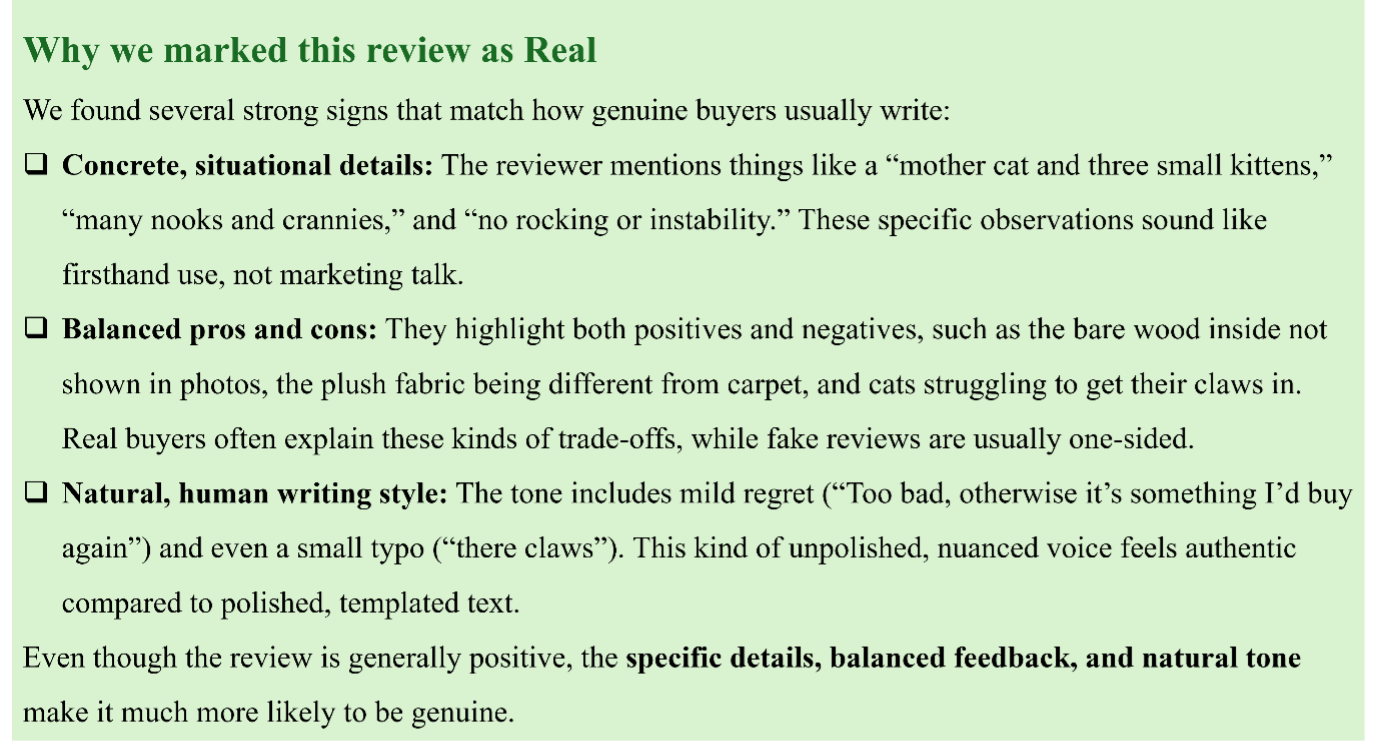}
    }

    \subfloat[\textbf{Word-highlighting based explanation}\label{fig:R2word}]{
        \includegraphics[width=0.7\linewidth]{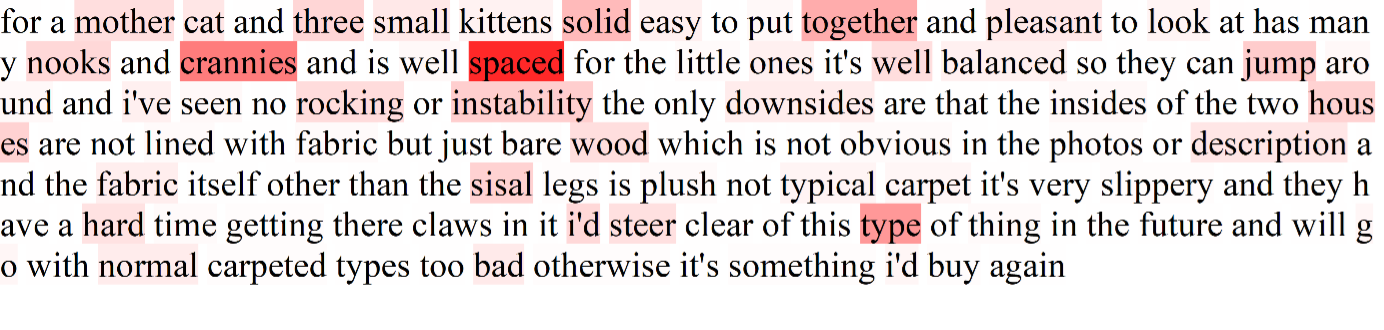}
    }

    \caption{Explanations generated by both systems for Review 2}
    \label{fig:R2Explanations}
\end{figure}

\newpage
\subsection{Review 3}
\noindent\textbf{Scenario: Electronics (Laptop Case with Keyboard Cover):} You’ve started taking your laptop to classes and want protection from scratches while also reducing noisy typing during lectures. Before buying an affordable laptop case with a keyboard cover, you check reviews to see if it meets those needs.\\

\noindent\textbf{Customer Review:} ``Overall, I really like this case. The rubberized outside feels smooth and sturdy and the pattern is nice, although not glossy as others have mentioned, which is fine with me. It feels like the case will be strong enough to protect my laptop from superficial scratches and such, but I'm not sure that it would prevent more severe damage with major impact, grantedt I don't expect much more given the very affordable price of the product. The keyboard cover is thinner than I imagined, but to be fair, I've never had a keyboard cover for my laptop so my only comparison was my imagination. It also has some creases in it from the way it was folded and packaged, so hopefully those will work themselves out with time and use. One of the main reasons I purchased this case was for the keyboard cover so that it would reduce the clicking sound while I take notes in class and it definitely achieves that.``

\begin{figure}[!htb]
    \centering

    \subfloat[\textbf{Language based Explanation}\newline\newline\label{fig:R3Lang}]{
        \includegraphics[width=0.7\linewidth]{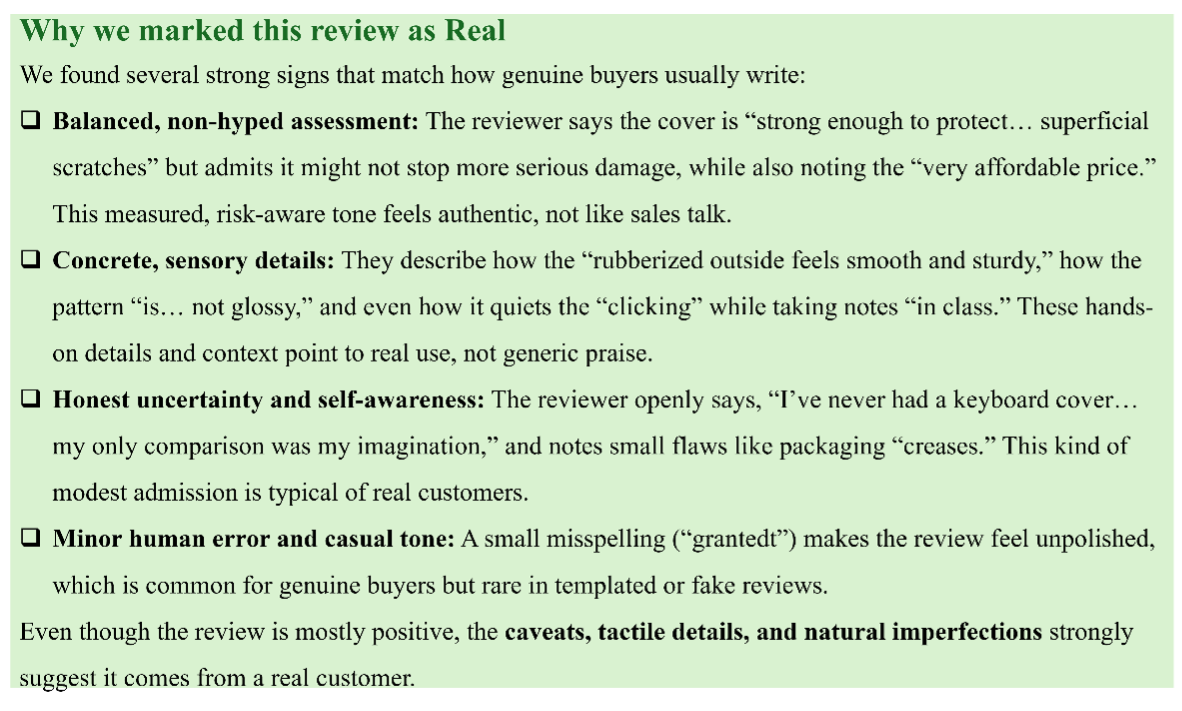}
    }

    \subfloat[\textbf{Word-highlighting based explanation}\label{fig:R3word}]{
        \includegraphics[width=0.7\linewidth]{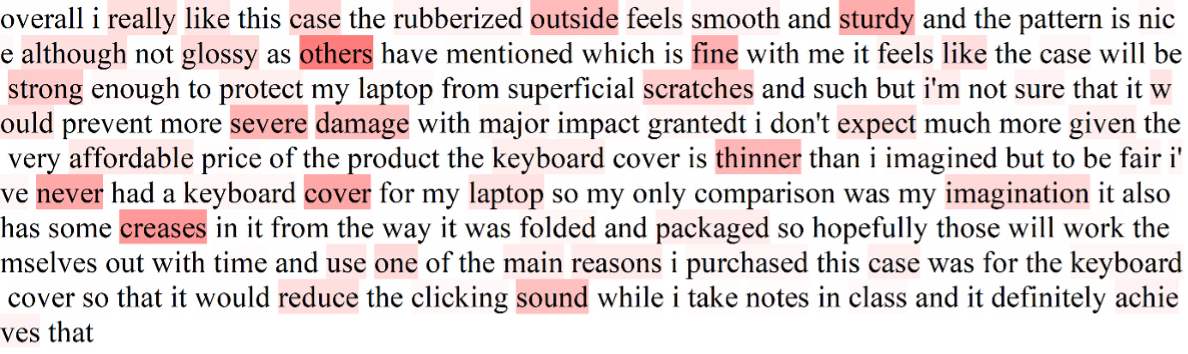}
    }

    \caption{Explanations generated by both systems for Review 3}
    \label{fig:R3Explanations}
\end{figure}

\newpage
\subsection{Review 4}

\noindent\textbf{Scenario: Toys and Games (Children Toy):} You’ve started taking your laptop to classes and want
protection from scratches while also reducing noisy typing during lectures. Before buying an affordable laptop case
with a keyboard cover, you check reviews to see if it meets those needs.

\noindent\textbf{Customer Review:} ``Its got everything I like about it, the materials are good, and the art is awesome.Great product. Great price! I am a huge fan of Tootlemania and this is by far my favorite. Highly recommend.I bought this for my daughter for Christmas. She loves it and it's great for her to keep her stuffed animals in the bath with her. It's made of very high quality and I highly recommend it for any girl who likes to play with toys.I got this for my son's birthday and he absolutely loves it! He plays with it every day.''

\begin{figure}[!htb]
    \centering

    \subfloat[\textbf{Language based Explanation}\newline\newline\label{fig:R4Lang}]{
        \includegraphics[width=0.7\linewidth]{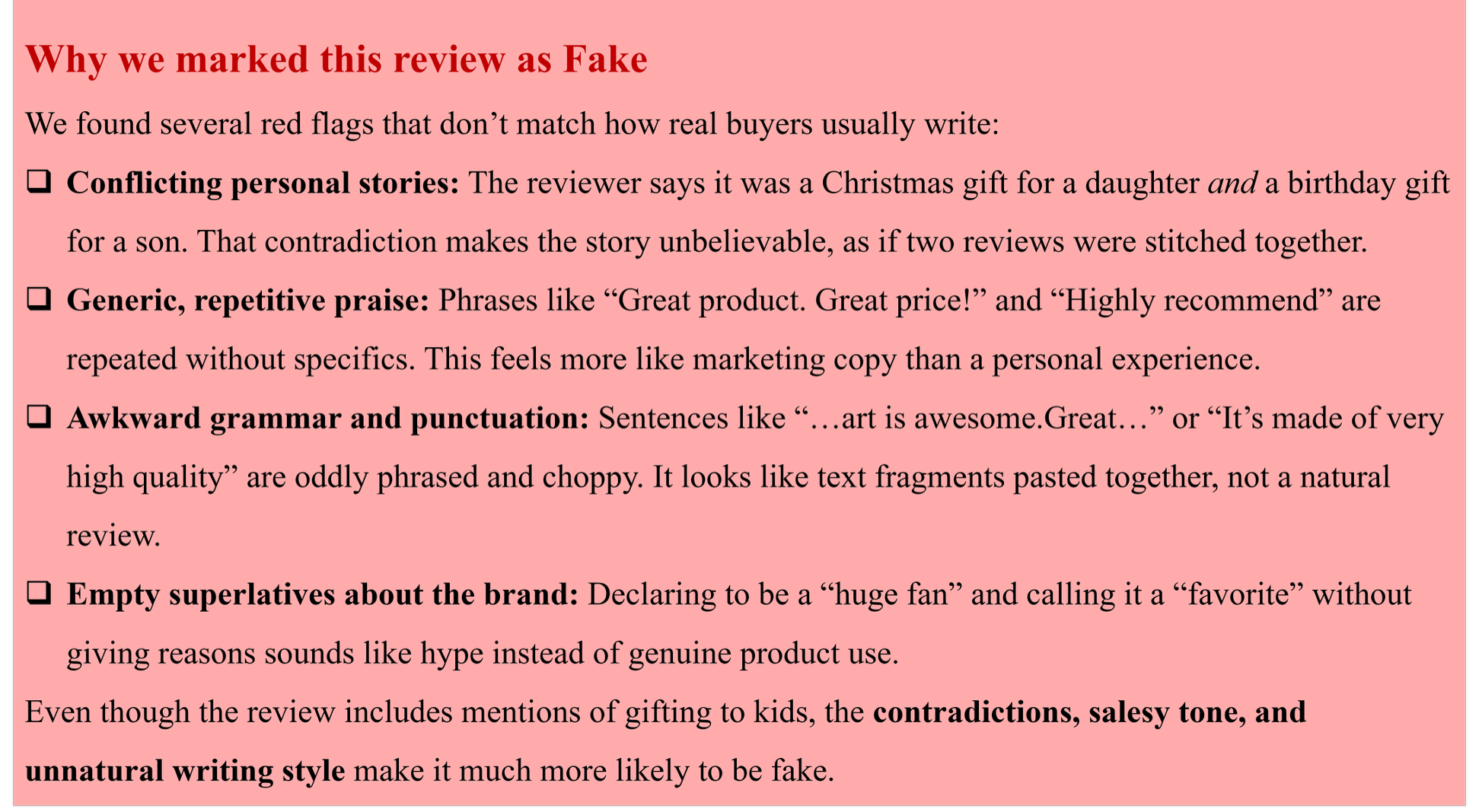}
    }

    \subfloat[\textbf{Word-highlighting based explanation}\label{fig:R4word}]{
        \includegraphics[width=0.7\linewidth]{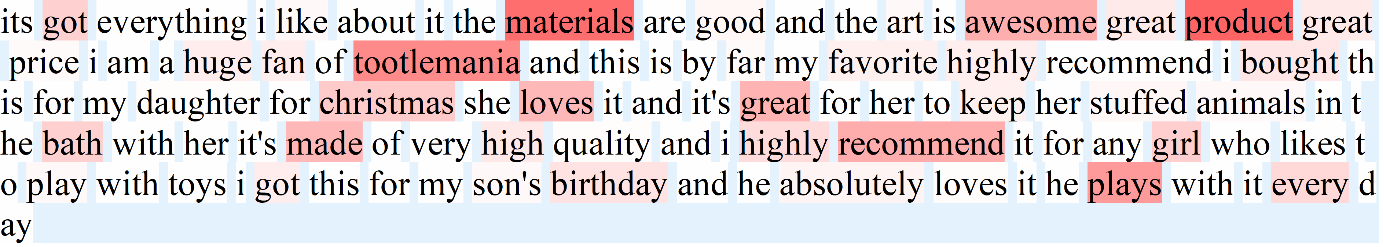}
    }

    \caption{Explanations generated by both systems for Review 4}
    \label{fig:R4Explanations}
\end{figure}

\newpage
\subsection{Review 5}

\noindent\textbf{Scenario: Kindle Store (E-Book):} You’re browsing for a new e-book to read during your daily commute. You come across this title but aren’t sure if the writing style and story will be engaging. To decide whether to buy, you look at reviews from other readers.

\noindent\textbf{Customer Review:} ``Valla is on a camping trip with his brothers and a few other people who have been through the past. The crew is having a hard time finding the words to describe what is going on. A little of the crew is trying to figure out what is going on. The crew is a bit confused because the crew is also engaged in a lot of other activities. They are trying to figure out what is going on and if they can't figure it out, they will have to go back to their campsites. They are trying to find out what is going on and if it is really a real problem. The next day, they have a meeting. The only problem is that the men are not at the meeting. They have to go back to their campsites and talk to each other. They are not going to be able to find the words to describe what is going on. The problem is that the men are not in town. They are in the town of Colosseum. They are not going to be able to find the words to describe what is going on. The problem is that the men are not talking to each other. They are not in town. They are in a bar. The men are in town. They are in the bar. They are in the bar. They are in a bar. They are in a bar. The men are in a bar. They are in the bar. They are in a bar.''

\begin{figure}[!htb]
    \centering

    \subfloat[\textbf{Language based Explanation}\newline\label{fig:R5Lang}]{
        \includegraphics[width=0.68\linewidth]{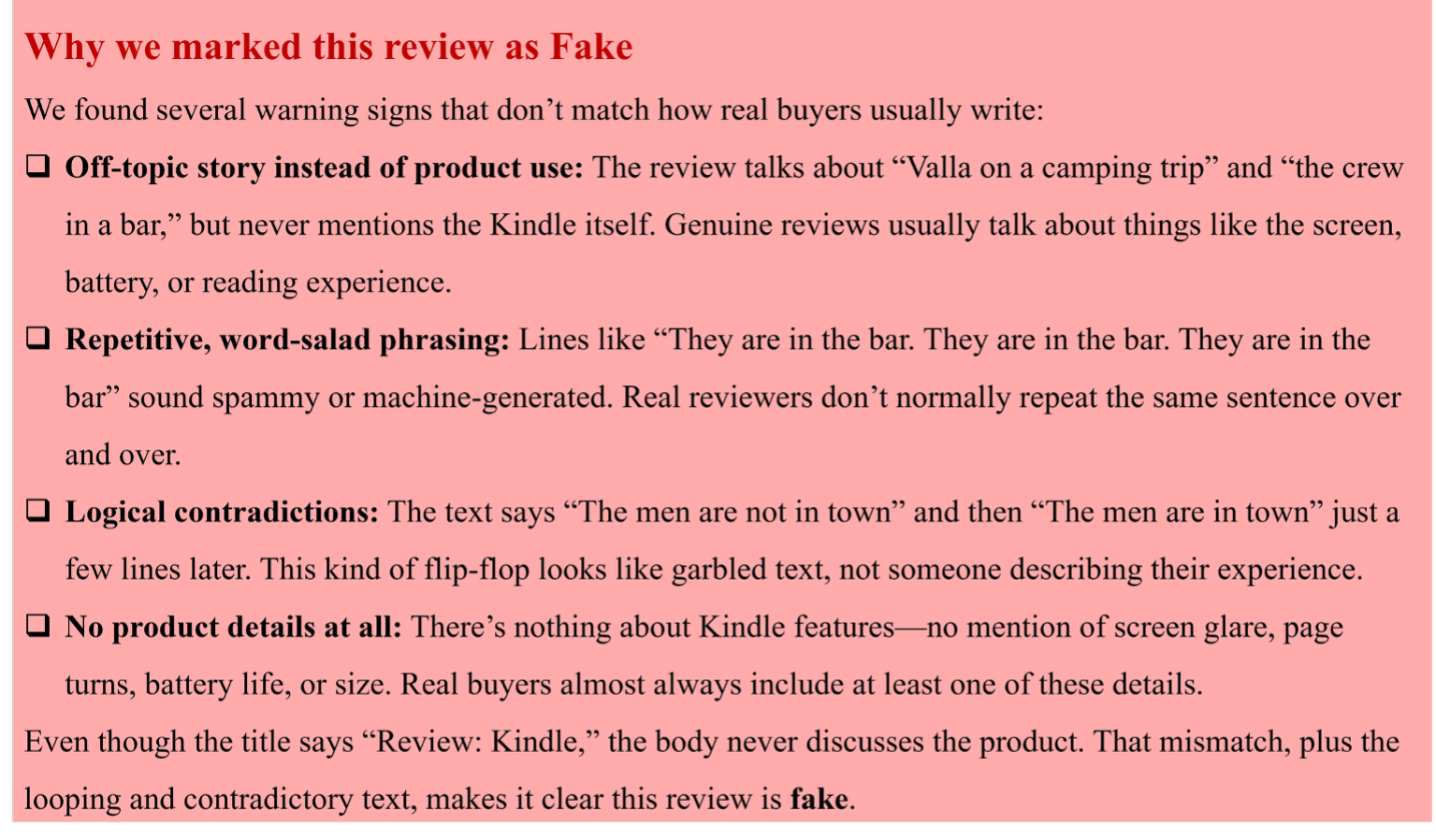}
    }

    \subfloat[\textbf{Word-highlighting based explanation}\label{fig:R5word}]{
        \includegraphics[width=0.68\linewidth]{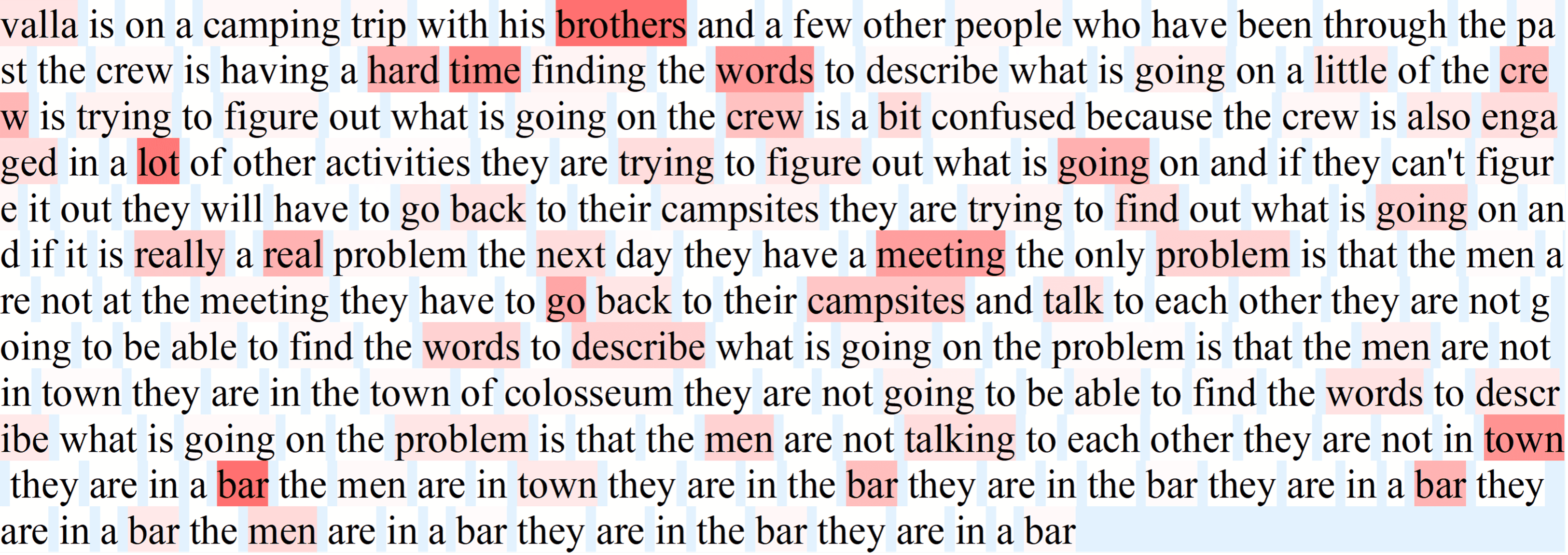}
    }

    \caption{Explanations generated by both systems for Review 5}
    \label{fig:R5Explanations}
\end{figure}

\newpage
\subsection{Review 6}
\noindent\textbf{Scenario: Toys and Games (Musical Toy):} You’re considering buying an educational toy for your young niece that combines music and play. It’s advertised as a piano, drum, and stacking game all in one. To avoid wasting money, you check reviews to see whether kids actually enjoy playing with it.

\noindent\textbf{Customer Review:} ``Disappointed with this one. It's a hybrid of sorts: it tries to be a song maker, a piano, and a stacker all in one. Unfortunately for small children, it's too bulky to really stack, and the drum and tambourine are too difficult to remove. The piano plays pre-set songs, or tracks of songs, rather than notes, so children have to listen thru before they engage again. Would have preferred individual sounds as (at least) an option. Because of the way the sound settings work, and the ease in which the cymbal can activate, ours essentially plays one song over and over again. We were definitely hoping for more, although it's not without its charms at times. ''

\begin{figure}[!htb]
    \centering

    \subfloat[\textbf{Language based Explanation}\newline\newline\label{fig:R6Lang}]{
        \includegraphics[width=0.7\linewidth]{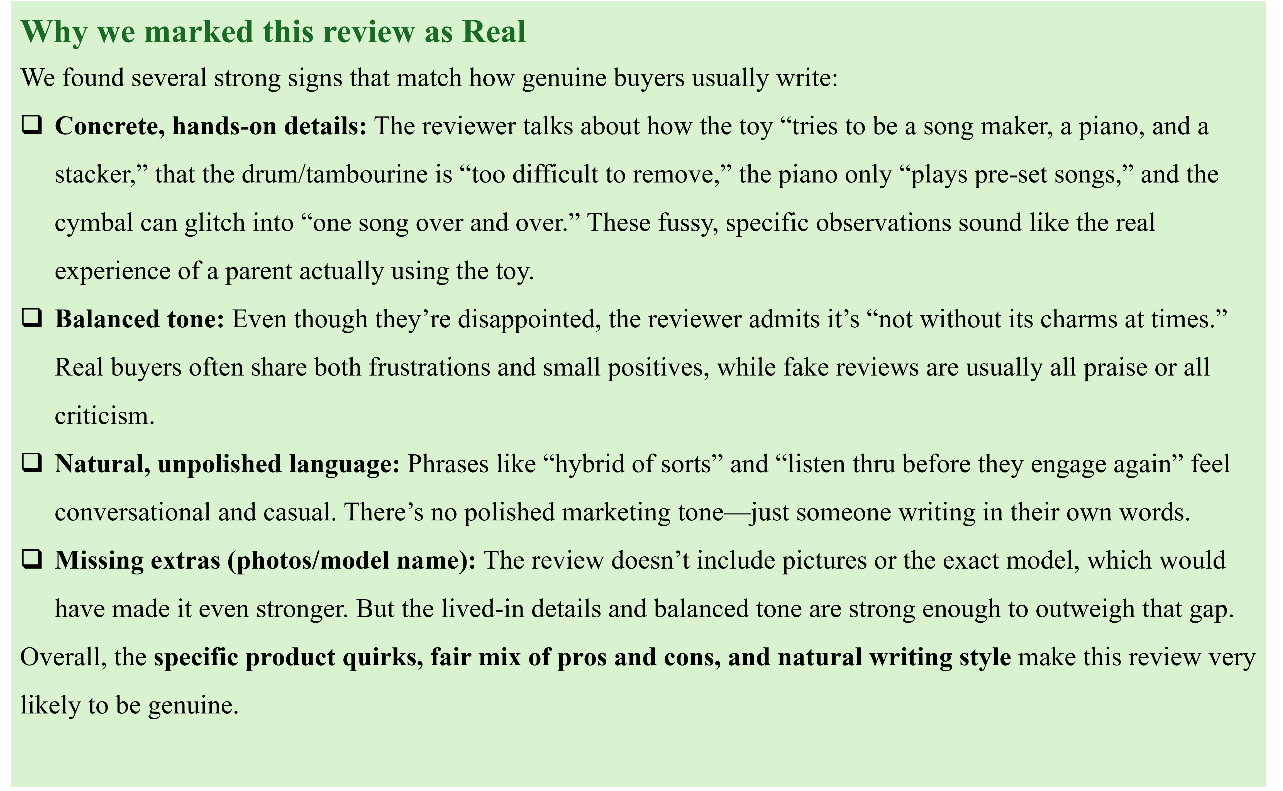}
    }

    \subfloat[\textbf{Word-highlighting based explanation}\label{fig:R6word}]{
        \includegraphics[width=0.7\linewidth]{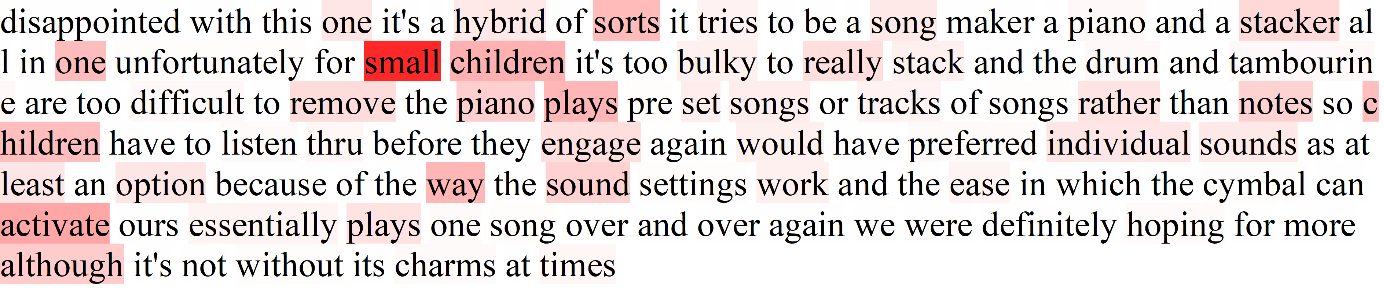}
    }

    \caption{Explanations generated by both systems for Review 6}
    \label{fig:R6Explanations}
\end{figure}
\end{document}